\documentclass[longauth]{aa}

\usepackage{graphicx}
\usepackage{natbib}
\usepackage{scalerel}
\usepackage{tabularx}
\usepackage[table]{xcolor}
\usepackage{siunitx}
\usepackage{lastpage}

\defcitealias{Q1-SP031}{E26}
\defcitealias{Q1-Laloux}{L26}
\defcitealias{Bisigello2025}{B25}
\DeclareSIUnit{\jansky}{Jy}
\DeclareSIUnit{\uJy}{\micro\jansky}

\usepackage{txfonts}
\usepackage[pdfencoding=auto,psdextra]{hyperref}
\hypersetup{
    colorlinks=true,
    linkcolor=blue,
    filecolor=magenta,      
    urlcolor=blue,
    citecolor=blue
}
\makeatletter
\renewcommand*\aa@pageof{, page \thepage{} of \pageref*{LastPage}}
\makeatother

\usepackage[utf8]{inputenc}
\usepackage[switch, modulo]{lineno}
\renewcommand{\linenumbers}[0]{}

\usepackage{euclid}
\usepackage[nolist,nohyperlinks]{acronym}

\begin{document}
\title{\Euclid: Quick Data Release (Q1) -- The evolution of the radio-loud AGN fraction-stellar mass relation to $z\sim3$ in the Euclid Deep Field North\thanks{This paper is published on behalf of the Euclid Consortium.}}    

\newcommand{\orcid}[1]{}   
\author{S.~Visser\orcid{0009-0004-4241-4911}\thanks{\email{visserscott@strw.leidenuniv.nl}}\inst{\ref{aff1}}
\and H.~J.~A.~Rottgering\orcid{0000-0001-8887-2257}\inst{\ref{aff1}}
\and L.~Wang\orcid{0000-0002-6736-9158}\inst{\ref{aff2},\ref{aff3}}
\and R.~Kondapally\orcid{0000-0001-6127-8151}\inst{\ref{aff4},\ref{aff5}}
\and S.~Tacchetto\orcid{0009-0008-0634-066X}\inst{\ref{aff6}}
\and B.~Laloux\orcid{0000-0001-9996-9732}\inst{\ref{aff7},\ref{aff8}}
\and L.~K.~Morabito\orcid{0000-0003-0487-6651}\inst{\ref{aff4},\ref{aff5}}
\and F.~Ricci\orcid{0000-0001-5742-5980}\inst{\ref{aff9},\ref{aff10}}
\and M.~Magliocchetti\orcid{0000-0001-9158-4838}\inst{\ref{aff11}}
\and A.~La~Marca\orcid{0000-0002-7217-5120}\inst{\ref{aff12},\ref{aff1}}
\and L.~Bisigello\orcid{0000-0003-0492-4924}\inst{\ref{aff13}}
\and A.~Viitanen\orcid{0000-0001-9383-786X}\inst{\ref{aff14},\ref{aff15},\ref{aff10}}
\and G.~J.~White\orcid{0000-0002-7126-691X}\inst{\ref{aff16},\ref{aff17}}
\and B.~Altieri\orcid{0000-0003-3936-0284}\inst{\ref{aff18}}
\and L.~Amendola\orcid{0000-0002-0835-233X}\inst{\ref{aff19},\ref{aff20}}
\and S.~Andreon\orcid{0000-0002-2041-8784}\inst{\ref{aff21}}
\and N.~Auricchio\orcid{0000-0003-4444-8651}\inst{\ref{aff22}}
\and C.~Baccigalupi\orcid{0000-0002-8211-1630}\inst{\ref{aff23},\ref{aff24},\ref{aff25},\ref{aff26}}
\and M.~Baldi\orcid{0000-0003-4145-1943}\inst{\ref{aff27},\ref{aff22},\ref{aff28}}
\and S.~Bardelli\orcid{0000-0002-8900-0298}\inst{\ref{aff22}}
\and P.~Battaglia\orcid{0000-0002-7337-5909}\inst{\ref{aff22}}
\and A.~Biviano\orcid{0000-0002-0857-0732}\inst{\ref{aff24},\ref{aff23}}
\and E.~Branchini\orcid{0000-0002-0808-6908}\inst{\ref{aff29},\ref{aff30},\ref{aff21}}
\and M.~Brescia\orcid{0000-0001-9506-5680}\inst{\ref{aff31},\ref{aff7}}
\and S.~Camera\orcid{0000-0003-3399-3574}\inst{\ref{aff32},\ref{aff33},\ref{aff34}}
\and V.~Capobianco\orcid{0000-0002-3309-7692}\inst{\ref{aff34}}
\and C.~Carbone\orcid{0000-0003-0125-3563}\inst{\ref{aff35}}
\and J.~Carretero\orcid{0000-0002-3130-0204}\inst{\ref{aff36},\ref{aff37}}
\and M.~Castellano\orcid{0000-0001-9875-8263}\inst{\ref{aff10}}
\and G.~Castignani\orcid{0000-0001-6831-0687}\inst{\ref{aff22}}
\and S.~Cavuoti\orcid{0000-0002-3787-4196}\inst{\ref{aff7},\ref{aff38}}
\and K.~C.~Chambers\orcid{0000-0001-6965-7789}\inst{\ref{aff39}}
\and A.~Cimatti\inst{\ref{aff40}}
\and C.~Colodro-Conde\inst{\ref{aff41}}
\and G.~Congedo\orcid{0000-0003-2508-0046}\inst{\ref{aff42}}
\and C.~J.~Conselice\orcid{0000-0003-1949-7638}\inst{\ref{aff43}}
\and L.~Conversi\orcid{0000-0002-6710-8476}\inst{\ref{aff44},\ref{aff18}}
\and Y.~Copin\orcid{0000-0002-5317-7518}\inst{\ref{aff45}}
\and F.~Courbin\orcid{0000-0003-0758-6510}\inst{\ref{aff46},\ref{aff47},\ref{aff48}}
\and H.~M.~Courtois\orcid{0000-0003-0509-1776}\inst{\ref{aff49}}
\and M.~Cropper\orcid{0000-0003-4571-9468}\inst{\ref{aff50}}
\and H.~Degaudenzi\orcid{0000-0002-5887-6799}\inst{\ref{aff14}}
\and G.~De~Lucia\orcid{0000-0002-6220-9104}\inst{\ref{aff24}}
\and H.~Dole\orcid{0000-0002-9767-3839}\inst{\ref{aff51}}
\and F.~Dubath\orcid{0000-0002-6533-2810}\inst{\ref{aff14}}
\and X.~Dupac\inst{\ref{aff18}}
\and S.~Dusini\orcid{0000-0002-1128-0664}\inst{\ref{aff52}}
\and M.~Fabricius\orcid{0000-0002-7025-6058}\inst{\ref{aff8},\ref{aff53}}
\and M.~Farina\orcid{0000-0002-3089-7846}\inst{\ref{aff11}}
\and R.~Farinelli\inst{\ref{aff22}}
\and F.~Faustini\orcid{0000-0001-6274-5145}\inst{\ref{aff10}}
\and S.~Ferriol\inst{\ref{aff45}}
\and M.~Frailis\orcid{0000-0002-7400-2135}\inst{\ref{aff24}}
\and E.~Franceschi\orcid{0000-0002-0585-6591}\inst{\ref{aff22}}
\and M.~Fumana\orcid{0000-0001-6787-5950}\inst{\ref{aff35}}
\and L.~Gabarra\orcid{0000-0002-8486-8856}\inst{\ref{aff54}}
\and S.~Galeotta\orcid{0000-0002-3748-5115}\inst{\ref{aff24}}
\and K.~George\orcid{0000-0002-1734-8455}\inst{\ref{aff55}}
\and B.~Gillis\orcid{0000-0002-4478-1270}\inst{\ref{aff42}}
\and C.~Giocoli\orcid{0000-0002-9590-7961}\inst{\ref{aff22},\ref{aff28}}
\and J.~Gracia-Carpio\orcid{0000-0003-4689-3134}\inst{\ref{aff8}}
\and A.~Grazian\orcid{0000-0002-5688-0663}\inst{\ref{aff13}}
\and F.~Grupp\inst{\ref{aff8},\ref{aff53}}
\and S.~Gwyn\orcid{0000-0001-8221-8406}\inst{\ref{aff56}}
\and W.~G.~Hartley\orcid{0000-0002-4563-5101}\inst{\ref{aff14}}
\and S.~V.~H.~Haugan\orcid{0000-0001-9648-7260}\inst{\ref{aff57}}
\and S.~Hemmati\orcid{0000-0003-2226-5395}\inst{\ref{aff58}}
\and H.~Hoekstra\orcid{0000-0002-0641-3231}\inst{\ref{aff1}}
\and W.~Holmes\orcid{0009-0007-8554-4646}\inst{\ref{aff59}}
\and I.~M.~Hook\orcid{0000-0002-2960-978X}\inst{\ref{aff60}}
\and F.~Hormuth\inst{\ref{aff61}}
\and A.~Hornstrup\orcid{0000-0002-3363-0936}\inst{\ref{aff62},\ref{aff63}}
\and K.~Jahnke\orcid{0000-0003-3804-2137}\inst{\ref{aff64}}
\and M.~Jhabvala\inst{\ref{aff65}}
\and S.~Kermiche\orcid{0000-0002-0302-5735}\inst{\ref{aff66}}
\and A.~Kiessling\orcid{0000-0002-2590-1273}\inst{\ref{aff59}}
\and B.~Kubik\orcid{0009-0006-5823-4880}\inst{\ref{aff45}}
\and K.~Kuijken\orcid{0000-0002-3827-0175}\inst{\ref{aff1}}
\and M.~K\"ummel\orcid{0000-0003-2791-2117}\inst{\ref{aff53}}
\and M.~Kunz\orcid{0000-0002-3052-7394}\inst{\ref{aff67}}
\and H.~Kurki-Suonio\orcid{0000-0002-4618-3063}\inst{\ref{aff68},\ref{aff69}}
\and A.~M.~C.~Le~Brun\orcid{0000-0002-0936-4594}\inst{\ref{aff70}}
\and V.~Le~Brun\orcid{0000-0002-5027-1939}\inst{\ref{aff71}}
\and S.~Ligori\orcid{0000-0003-4172-4606}\inst{\ref{aff34}}
\and P.~B.~Lilje\orcid{0000-0003-4324-7794}\inst{\ref{aff57}}
\and V.~Lindholm\orcid{0000-0003-2317-5471}\inst{\ref{aff68},\ref{aff69}}
\and I.~Lloro\orcid{0000-0001-5966-1434}\inst{\ref{aff72}}
\and G.~Mainetti\orcid{0000-0003-2384-2377}\inst{\ref{aff73}}
\and O.~Mansutti\orcid{0000-0001-5758-4658}\inst{\ref{aff24}}
\and O.~Marggraf\orcid{0000-0001-7242-3852}\inst{\ref{aff74}}
\and M.~Martinelli\orcid{0000-0002-6943-7732}\inst{\ref{aff10},\ref{aff75}}
\and N.~Martinet\orcid{0000-0003-2786-7790}\inst{\ref{aff71}}
\and F.~Marulli\orcid{0000-0002-8850-0303}\inst{\ref{aff76},\ref{aff22},\ref{aff28}}
\and R.~J.~Massey\orcid{0000-0002-6085-3780}\inst{\ref{aff5}}
\and N.~Mauri\orcid{0000-0001-8196-1548}\inst{\ref{aff40},\ref{aff28}}
\and E.~Medinaceli\orcid{0000-0002-4040-7783}\inst{\ref{aff22}}
\and S.~Mei\orcid{0000-0002-2849-559X}\inst{\ref{aff77},\ref{aff78}}
\and M.~Meneghetti\orcid{0000-0003-1225-7084}\inst{\ref{aff22},\ref{aff28}}
\and E.~Merlin\orcid{0000-0001-6870-8900}\inst{\ref{aff13}}
\and G.~Meylan\orcid{0000-0001-6503-0209}\inst{\ref{aff79}}
\and J.~J.~Mohr\orcid{0000-0002-6875-2087}\inst{\ref{aff55}}
\and P.~Monaco\orcid{0000-0003-2083-7564}\inst{\ref{aff80},\ref{aff24},\ref{aff25},\ref{aff23}}
\and A.~Mora\orcid{0000-0002-1922-8529}\inst{\ref{aff81}}
\and M.~Moresco\orcid{0000-0002-7616-7136}\inst{\ref{aff76},\ref{aff22}}
\and C.~Moretti\orcid{0000-0003-3314-8936}\inst{\ref{aff24},\ref{aff23},\ref{aff25}}
\and L.~Moscardini\orcid{0000-0002-3473-6716}\inst{\ref{aff76},\ref{aff22},\ref{aff28}}
\and C.~Neissner\orcid{0000-0001-8524-4968}\inst{\ref{aff82},\ref{aff37}}
\and S.-M.~Niemi\orcid{0009-0005-0247-0086}\inst{\ref{aff12}}
\and C.~Padilla\orcid{0000-0001-7951-0166}\inst{\ref{aff82}}
\and S.~Paltani\orcid{0000-0002-8108-9179}\inst{\ref{aff14}}
\and F.~Pasian\orcid{0000-0002-4869-3227}\inst{\ref{aff24}}
\and W.~J.~Percival\orcid{0000-0002-0644-5727}\inst{\ref{aff83},\ref{aff84},\ref{aff85}}
\and V.~Pettorino\orcid{0000-0002-4203-9320}\inst{\ref{aff12}}
\and A.~Pezzotta\orcid{0000-0003-0726-2268}\inst{\ref{aff21}}
\and G.~Polenta\orcid{0000-0003-4067-9196}\inst{\ref{aff86}}
\and M.~Poncet\inst{\ref{aff87}}
\and L.~A.~Popa\inst{\ref{aff88}}
\and L.~Pozzetti\orcid{0000-0001-7085-0412}\inst{\ref{aff22}}
\and F.~Raison\orcid{0000-0002-7819-6918}\inst{\ref{aff8}}
\and A.~Renzi\orcid{0000-0001-9856-1970}\inst{\ref{aff89},\ref{aff52},\ref{aff22}}
\and J.~Rhodes\orcid{0000-0002-4485-8549}\inst{\ref{aff59}}
\and G.~Riccio\inst{\ref{aff7}}
\and I.~Risso\orcid{0000-0003-2525-7761}\inst{\ref{aff29},\ref{aff30},\ref{aff21}}
\and F.~Rizzo\orcid{0000-0002-9407-585X}\inst{\ref{aff24}}
\and E.~Romelli\orcid{0000-0003-3069-9222}\inst{\ref{aff24}}
\and M.~Roncarelli\orcid{0000-0001-9587-7822}\inst{\ref{aff22}}
\and B.~Rusholme\orcid{0000-0001-7648-4142}\inst{\ref{aff58}}
\and R.~Saglia\orcid{0000-0003-0378-7032}\inst{\ref{aff53},\ref{aff8}}
\and Z.~Sakr\orcid{0000-0002-4823-3757}\inst{\ref{aff90},\ref{aff91},\ref{aff92}}
\and D.~Sapone\orcid{0000-0001-7089-4503}\inst{\ref{aff93}}
\and B.~Sartoris\orcid{0000-0003-1337-5269}\inst{\ref{aff8},\ref{aff24}}
\and J.~A.~Schewtschenko\orcid{0000-0002-4913-6393}\inst{\ref{aff42}}
\and M.~Schirmer\orcid{0000-0003-2568-9994}\inst{\ref{aff64}}
\and P.~Schneider\orcid{0000-0001-8561-2679}\inst{\ref{aff74}}
\and T.~Schrabback\orcid{0000-0002-6987-7834}\inst{\ref{aff6}}
\and A.~Secroun\orcid{0000-0003-0505-3710}\inst{\ref{aff66}}
\and E.~Sihvola\orcid{0000-0003-1804-7715}\inst{\ref{aff15}}
\and P.~Simon\inst{\ref{aff74}}
\and C.~Sirignano\orcid{0000-0002-0995-7146}\inst{\ref{aff89},\ref{aff52}}
\and G.~Sirri\orcid{0000-0003-2626-2853}\inst{\ref{aff28}}
\and P.~Tallada-Cresp\'{i}\orcid{0000-0002-1336-8328}\inst{\ref{aff36},\ref{aff37}}
\and A.~N.~Taylor\inst{\ref{aff42}}
\and I.~Tereno\orcid{0000-0002-4537-6218}\inst{\ref{aff94},\ref{aff95}}
\and N.~Tessore\orcid{0000-0002-9696-7931}\inst{\ref{aff50}}
\and S.~Toft\orcid{0000-0003-3631-7176}\inst{\ref{aff96},\ref{aff97}}
\and R.~Toledo-Moreo\orcid{0000-0002-2997-4859}\inst{\ref{aff98},\ref{aff99}}
\and F.~Torradeflot\orcid{0000-0003-1160-1517}\inst{\ref{aff37},\ref{aff36}}
\and I.~Tutusaus\orcid{0000-0002-3199-0399}\inst{\ref{aff100},\ref{aff101},\ref{aff91}}
\and J.~Valiviita\orcid{0000-0001-6225-3693}\inst{\ref{aff68},\ref{aff69}}
\and T.~Vassallo\orcid{0000-0001-6512-6358}\inst{\ref{aff24},\ref{aff55}}
\and A.~Veropalumbo\orcid{0000-0003-2387-1194}\inst{\ref{aff21},\ref{aff30},\ref{aff29}}
\and Y.~Wang\orcid{0000-0002-4749-2984}\inst{\ref{aff58}}
\and J.~Weller\orcid{0000-0002-8282-2010}\inst{\ref{aff53},\ref{aff8}}
\and A.~Zacchei\orcid{0000-0003-0396-1192}\inst{\ref{aff24},\ref{aff23}}
\and G.~Zamorani\orcid{0000-0002-2318-301X}\inst{\ref{aff22}}
\and F.~M.~Zerbi\orcid{0000-0002-9996-973X}\inst{\ref{aff21}}
\and E.~Zucca\orcid{0000-0002-5845-8132}\inst{\ref{aff22}}
\and C.~M.~Gutierrez\orcid{0000-0001-7854-783X}\inst{\ref{aff41},\ref{aff102}}
\and K.~Tanidis\orcid{0000-0001-9843-5130}\inst{\ref{aff103}}
\and D.~Scott\orcid{0000-0002-6878-9840}\inst{\ref{aff104}}}
										   
\institute{Leiden Observatory, Leiden University, Einsteinweg 55, 2333 CC Leiden, The Netherlands\label{aff1}
\and
SRON Netherlands Institute for Space Research, Landleven 12, 9747 AD, Groningen, The Netherlands\label{aff2}
\and
Kapteyn Astronomical Institute, University of Groningen, PO Box 800, 9700 AV Groningen, The Netherlands\label{aff3}
\and
Department of Physics, Centre for Extragalactic Astronomy, Durham University, South Road, Durham, DH1 3LE, UK\label{aff4}
\and
Department of Physics, Institute for Computational Cosmology, Durham University, South Road, Durham, DH1 3LE, UK\label{aff5}
\and
Universit\"at Innsbruck, Institut f\"ur Astro- und Teilchenphysik, Technikerstr. 25/8, 6020 Innsbruck, Austria\label{aff6}
\and
INAF-Osservatorio Astronomico di Capodimonte, Via Moiariello 16, 80131 Napoli, Italy\label{aff7}
\and
Max Planck Institute for Extraterrestrial Physics, Giessenbachstr. 1, 85748 Garching, Germany\label{aff8}
\and
Department of Mathematics and Physics, Roma Tre University, Via della Vasca Navale 84, 00146 Rome, Italy\label{aff9}
\and
INAF-Osservatorio Astronomico di Roma, Via Frascati 33, 00078 Monteporzio Catone, Italy\label{aff10}
\and
INAF-Istituto di Astrofisica e Planetologia Spaziali, via del Fosso del Cavaliere, 100, 00100 Roma, Italy\label{aff11}
\and
European Space Agency/ESTEC, Keplerlaan 1, 2201 AZ Noordwijk, The Netherlands\label{aff12}
\and
INAF-Osservatorio Astronomico di Padova, Via dell'Osservatorio 5, 35122 Padova, Italy\label{aff13}
\and
Department of Astronomy, University of Geneva, ch. d'Ecogia 16, 1290 Versoix, Switzerland\label{aff14}
\and
Department of Physics and Helsinki Institute of Physics, Gustaf H\"allstr\"omin katu 2, University of Helsinki, 00014 Helsinki, Finland\label{aff15}
\and
School of Physical Sciences, The Open University, Milton Keynes, MK7 6AA, UK\label{aff16}
\and
RAL Space, Rutherford Appleton Laboratory, STFC, UKRI, Harwell Campus, Oxfordshire, OX11 0QX, UK\label{aff17}
\and
ESAC/ESA, Camino Bajo del Castillo, s/n., Urb. Villafranca del Castillo, 28692 Villanueva de la Ca\~nada, Madrid, Spain\label{aff18}
\and
Institut f\"ur Theoretische Physik, University of Heidelberg, Philosophenweg 16, 69120 Heidelberg, Germany\label{aff19}
\and
New York University Abu Dhabi, PO Box 129188, Abu Dhabi, UAE, and Center for Astrophysics and Space Science (CASS), New York University Abu Dhabi, UAE\label{aff20}
\and
INAF-Osservatorio Astronomico di Brera, Via Brera 28, 20122 Milano, Italy\label{aff21}
\and
INAF-Osservatorio di Astrofisica e Scienza dello Spazio di Bologna, Via Piero Gobetti 93/3, 40129 Bologna, Italy\label{aff22}
\and
IFPU, Institute for Fundamental Physics of the Universe, via Beirut 2, 34151 Trieste, Italy\label{aff23}
\and
INAF-Osservatorio Astronomico di Trieste, Via G. B. Tiepolo 11, 34143 Trieste, Italy\label{aff24}
\and
INFN, Sezione di Trieste, Via Valerio 2, 34127 Trieste TS, Italy\label{aff25}
\and
SISSA, International School for Advanced Studies, Via Bonomea 265, 34136 Trieste TS, Italy\label{aff26}
\and
Dipartimento di Fisica e Astronomia, Universit\`a di Bologna, Via Gobetti 93/2, 40129 Bologna, Italy\label{aff27}
\and
INFN-Sezione di Bologna, Viale Berti Pichat 6/2, 40127 Bologna, Italy\label{aff28}
\and
Dipartimento di Fisica, Universit\`a di Genova, Via Dodecaneso 33, 16146, Genova, Italy\label{aff29}
\and
INFN-Sezione di Genova, Via Dodecaneso 33, 16146, Genova, Italy\label{aff30}
\and
Department of Physics "E. Pancini", University Federico II, Via Cinthia 6, 80126, Napoli, Italy\label{aff31}
\and
Dipartimento di Fisica, Universit\`a degli Studi di Torino, Via P. Giuria 1, 10125 Torino, Italy\label{aff32}
\and
INFN-Sezione di Torino, Via P. Giuria 1, 10125 Torino, Italy\label{aff33}
\and
INAF-Osservatorio Astrofisico di Torino, Via Osservatorio 20, 10025 Pino Torinese (TO), Italy\label{aff34}
\and
INAF-IASF Milano, Via Alfonso Corti 12, 20133 Milano, Italy\label{aff35}
\and
Centro de Investigaciones Energ\'eticas, Medioambientales y Tecnol\'ogicas (CIEMAT), Avenida Complutense 40, 28040 Madrid, Spain\label{aff36}
\and
Port d'Informaci\'{o} Cient\'{i}fica, Campus UAB, C. Albareda s/n, 08193 Bellaterra (Barcelona), Spain\label{aff37}
\and
INFN -- Sezione di Napoli, Via Cinthia 6, 80126, Napoli, Italy\label{aff38}
\and
Institute for Astronomy, University of Hawaii, 2680 Woodlawn Drive, Honolulu, HI 96822, USA\label{aff39}
\and
Dipartimento di Fisica e Astronomia "Augusto Righi" - Alma Mater Studiorum Universit\`a di Bologna, Viale Berti Pichat 6/2, 40127 Bologna, Italy\label{aff40}
\and
Instituto de Astrof\'{\i}sica de Canarias, E-38205 La Laguna, Tenerife, Spain\label{aff41}
\and
Institute for Astronomy, University of Edinburgh, Royal Observatory, Blackford Hill, Edinburgh EH9 3HJ, UK\label{aff42}
\and
Jodrell Bank Centre for Astrophysics, Department of Physics and Astronomy, University of Manchester, Oxford Road, Manchester M13 9PL, UK\label{aff43}
\and
European Space Agency/ESRIN, Largo Galileo Galilei 1, 00044 Frascati, Roma, Italy\label{aff44}
\and
Universit\'e Lyon 1, CNRS/IN2P3, IP2I Lyon, Villeurbanne, France\label{aff45}
\and
Institut de Ci\`{e}ncies del Cosmos (ICCUB), Universitat de Barcelona (IEEC-UB), Mart\'{i} i Franqu\`{e}s 1, 08028 Barcelona, Spain\label{aff46}
\and
Instituci\'o Catalana de Recerca i Estudis Avan\c{c}ats (ICREA), Passeig de Llu\'{\i}s Companys 23, 08010 Barcelona, Spain\label{aff47}
\and
Institut de Ciencies de l'Espai (IEEC-CSIC), Campus UAB, Carrer de Can Magrans, s/n Cerdanyola del Vall\'es, 08193 Barcelona, Spain\label{aff48}
\and
Universit\'e Lyon 1, CNRS/IN2P3, IUF, IP2I Lyon, Villeurbanne, France \label{aff49}
\and
Mullard Space Science Laboratory, University College London, Holmbury St Mary, Dorking, Surrey RH5 6NT, UK\label{aff50}
\and
Universit\'e Paris-Saclay, CNRS, Institut d'astrophysique spatiale, 91405, Orsay, France\label{aff51}
\and
INFN-Padova, Via Marzolo 8, 35131 Padova, Italy\label{aff52}
\and
Universit\"ats-Sternwarte M\"unchen, Fakult\"at f\"ur Physik, Ludwig-Maximilians-Universit\"at M\"unchen, Scheinerstr.~1, 81679 M\"unchen, Germany\label{aff53}
\and
Department of Physics, University of Oxford, Keble Road, Oxford OX1 3RH, UK\label{aff54}
\and
University Observatory, LMU Faculty of Physics, Scheinerstr.~1, 81679 Munich, Germany\label{aff55}
\and
National Research Council, Herzberg Astronomy and Astrophysics Research Centre, 5071 W. Saanich Rd. Victoria, BC, V9E 2E7, Canada\label{aff56}
\and
Institute of Theoretical Astrophysics, University of Oslo, P.O. Box 1029 Blindern, 0315 Oslo, Norway\label{aff57}
\and
Caltech/IPAC, 1200 E. California Blvd., Pasadena, CA 91125, USA\label{aff58}
\and
Jet Propulsion Laboratory, California Institute of Technology, 4800 Oak Grove Drive, Pasadena, CA, 91109, USA\label{aff59}
\and
School of Physics and Astronomy, Lancaster University, Lancaster, LA1 4YB, UK\label{aff60}
\and
Felix Hormuth Engineering, Goethestr. 17, 69181 Leimen, Germany\label{aff61}
\and
Technical University of Denmark, Elektrovej 327, 2800 Kgs. Lyngby, Denmark\label{aff62}
\and
Cosmic Dawn Center (DAWN), Denmark\label{aff63}
\and
Max-Planck-Institut f\"ur Astronomie, K\"onigstuhl 17, 69117 Heidelberg, Germany\label{aff64}
\and
NASA Goddard Space Flight Center, Greenbelt, MD 20771, USA\label{aff65}
\and
Aix-Marseille Universit\'e, CNRS/IN2P3, CPPM, Marseille, France\label{aff66}
\and
Universit\'e de Gen\`eve, D\'epartement de Physique Th\'eorique and Centre for Astroparticle Physics, 24 quai Ernest-Ansermet, CH-1211 Gen\`eve 4, Switzerland\label{aff67}
\and
Department of Physics, P.O. Box 64, University of Helsinki, 00014 Helsinki, Finland\label{aff68}
\and
Helsinki Institute of Physics, Gustaf H{\"a}llstr{\"o}min katu 2, University of Helsinki, 00014 Helsinki, Finland\label{aff69}
\and
Laboratoire d'etude de l'Univers et des phenomenes eXtremes, Observatoire de Paris, Universit\'e PSL, Sorbonne Universit\'e, CNRS, 92190 Meudon, France\label{aff70}
\and
Aix-Marseille Universit\'e, CNRS, CNES, LAM, Marseille, France\label{aff71}
\and
SKAO, Jodrell Bank, Lower Withington, Macclesfield SK11 9FT, UK\label{aff72}
\and
Centre de Calcul de l'IN2P3/CNRS, 21 avenue Pierre de Coubertin 69627 Villeurbanne Cedex, France\label{aff73}
\and
Universit\"at Bonn, Argelander-Institut f\"ur Astronomie, Auf dem H\"ugel 71, 53121 Bonn, Germany\label{aff74}
\and
INFN-Sezione di Roma, Piazzale Aldo Moro, 2 - c/o Dipartimento di Fisica, Edificio G. Marconi, 00185 Roma, Italy\label{aff75}
\and
Dipartimento di Fisica e Astronomia "Augusto Righi" - Alma Mater Studiorum Universit\`a di Bologna, via Piero Gobetti 93/2, 40129 Bologna, Italy\label{aff76}
\and
Universit\'e Paris Cit\'e, CNRS, Astroparticule et Cosmologie, 75013 Paris, France\label{aff77}
\and
CNRS-UCB International Research Laboratory, Centre Pierre Bin\'etruy, IRL2007, CPB-IN2P3, Berkeley, USA\label{aff78}
\and
Institute of Physics, Laboratory of Astrophysics, Ecole Polytechnique F\'ed\'erale de Lausanne (EPFL), Observatoire de Sauverny, 1290 Versoix, Switzerland\label{aff79}
\and
Dipartimento di Fisica - Sezione di Astronomia, Universit\`a di Trieste, Via Tiepolo 11, 34131 Trieste, Italy\label{aff80}
\and
Telespazio UK S.L. for European Space Agency (ESA), Camino bajo del Castillo, s/n, Urbanizacion Villafranca del Castillo, Villanueva de la Ca\~nada, 28692 Madrid, Spain\label{aff81}
\and
Institut de F\'{i}sica d'Altes Energies (IFAE), The Barcelona Institute of Science and Technology, Campus UAB, 08193 Bellaterra (Barcelona), Spain\label{aff82}
\and
Waterloo Centre for Astrophysics, University of Waterloo, Waterloo, Ontario N2L 3G1, Canada\label{aff83}
\and
Department of Physics and Astronomy, University of Waterloo, Waterloo, Ontario N2L 3G1, Canada\label{aff84}
\and
Perimeter Institute for Theoretical Physics, Waterloo, Ontario N2L 2Y5, Canada\label{aff85}
\and
Space Science Data Center, Italian Space Agency, via del Politecnico snc, 00133 Roma, Italy\label{aff86}
\and
Centre National d'Etudes Spatiales -- Centre spatial de Toulouse, 18 avenue Edouard Belin, 31401 Toulouse Cedex 9, France\label{aff87}
\and
Institute of Space Science, Str. Atomistilor, nr. 409 M\u{a}gurele, Ilfov, 077125, Romania\label{aff88}
\and
Dipartimento di Fisica e Astronomia "G. Galilei", Universit\`a di Padova, Via Marzolo 8, 35131 Padova, Italy\label{aff89}
\and
Instituto de F\'isica Te\'orica UAM-CSIC, Campus de Cantoblanco, 28049 Madrid, Spain\label{aff90}
\and
Institut de Recherche en Astrophysique et Plan\'etologie (IRAP), Universit\'e de Toulouse, CNRS, UPS, CNES, 14 Av. Edouard Belin, 31400 Toulouse, France\label{aff91}
\and
Universit\'e St Joseph; Faculty of Sciences, Beirut, Lebanon\label{aff92}
\and
Departamento de F\'isica, FCFM, Universidad de Chile, Blanco Encalada 2008, Santiago, Chile\label{aff93}
\and
Departamento de F\'isica, Faculdade de Ci\^encias, Universidade de Lisboa, Edif\'icio C8, Campo Grande, PT1749-016 Lisboa, Portugal\label{aff94}
\and
Instituto de Astrof\'isica e Ci\^encias do Espa\c{c}o, Faculdade de Ci\^encias, Universidade de Lisboa, Tapada da Ajuda, 1349-018 Lisboa, Portugal\label{aff95}
\and
Cosmic Dawn Center (DAWN)\label{aff96}
\and
Niels Bohr Institute, University of Copenhagen, Jagtvej 128, 2200 Copenhagen, Denmark\label{aff97}
\and
Universidad Polit\'ecnica de Cartagena, Departamento de Electr\'onica y Tecnolog\'ia de Computadoras,  Plaza del Hospital 1, 30202 Cartagena, Spain\label{aff98}
\and
European University of Technology EUt+, European Union\label{aff99}
\and
Institute of Space Sciences (ICE, CSIC), Campus UAB, Carrer de Can Magrans, s/n, 08193 Barcelona, Spain\label{aff100}
\and
Institut d'Estudis Espacials de Catalunya (IEEC),  Edifici RDIT, Campus UPC, 08860 Castelldefels, Barcelona, Spain\label{aff101}
\and
Universidad de La Laguna, Dpto. Astrof\'\i sica, E-38206 La Laguna, Tenerife, Spain\label{aff102}
\and
Center for Astrophysics and Cosmology, University of Nova Gorica, Nova Gorica, Slovenia\label{aff103}
\and
Department of Physics and Astronomy, University of British Columbia, Vancouver, BC V6T 1Z1, Canada\label{aff104}}    

\acrodef{agn}[AGN]{active galactic nucleus}
\acrodef{smbh}[SMBH]{supermassive black hole}
\acrodef{adaf}[ADAF]{advection-dominated accretion flow}
\acrodef{sfr}[SFR]{star-formation rate}
\acrodef{snr}[S/N]{signal-to-noise ratio}
\acrodef{ra}[RA]{right ascension}
\acrodef{dec}[Dec]{declination}
\acrodef{ssfr}[sSFR]{specific SFR}
\acrodef{nn}[NN]{nearest-neighbour}
\acrodef{odr}[ODR]{orthogonal distance regression}
\acrodef{sfg}[SFGs]{star-forming galaxies}
\acrodef{sed}[SED]{spectral energy distribution}
\acrodef{desi}[DESI]{Dark Energy Spectroscopic Instrument}
\acrodef{fwhm}[FWHM]{full width at half maximum}
\acrodef{pdf}[PDF]{probability distribution function}
\acrodef{cigale}[\texttt{CIGALE}]{Code Investigating GALaxy Emission}
\acrodef{nmad}[NMAD]{normalised median absolute deviation}
\acrodef{nir}[NIR]{near-infrared}
\acrodef{lofar}[LOFAR]{LOw Frequency ARray}
\acrodef{lotss}[LoTSS]{LOFAR Two Meter Sky Survey}
\acrodef{lr}[LR]{likelihood ratio}
\acrodef{vlass}[VLASS]{Very Large Array Sky Survey}
\acrodef{vla}[VLA]{Very Large Array}
\acrodef{rlagn}[RLAGN]{radio-loud AGN}
\acrodef{herg}[HERG]{high-excitation radio galaxy}
\acrodef{lerg}[LERG]{low-excitation radio galaxy}
\acrodef{hba}[HBA]{High Band Antenna}
\acrodef{pybdsf}[\texttt{PyBDSF}]{Python Blob Detection and Source Finder}
\acrodef{ilt}[ILT]{International LOFAR Telescope}
\acrodef{nvss}[NVSS]{NRAO Very Large Array Sky Survey}
\acrodef{hetdex}[HETDEX]{Hobby-Eberly Telescope Dark Energy Experiment}
\acrodef{rms}[RMS]{root mean square}
\acrodef{mightee}[MIGHTEE]{MeerKAT International GHz Tiered Extragalactic Exploration}
\acrodef{first}[FIRST]{Faint Images of the Radio Sky at Twenty-Centimeters}
\acrodef{edfn}[EDF-N]{Euclid Deep Field North}
\acrodef{edf}[EDFs]{Euclid Deep Fields}
\acrodef{nisp}[NISP]{Near Infrared Spectro- and Photometer}
\acrodef{irac}[IRAC]{Infrared Array Camera}
\acrodef{sdss}[SDSS]{Sloan Digital Sky Survey}
\acrodef{wht}[WHT]{William Herschel Telescope}
\acrodef{weave}[WEAVE]{WHT Enhanced Area Velocity Explorer}

\abstract{
The incidence of radio-loud active galactic nuclei (RLAGN) and their energetic output via AGN feedback are crucial components in understanding the evolution of galaxies and the large-scale structure of the Universe. In the local Universe, the incidence of RLAGN has been shown to correlate strongly with host galaxy properties, particularly stellar mass. Using the large-area survey capabilities provided by \Euclid and the LOw Frequency ARray (LOFAR), we present the first study of the RLAGN incidence--stellar mass relation to $z = 3.0$. We perform a stringent analysis taking into account cross-matching reliability and stellar-mass incompleteness. In addition, we test the robustness of our results against AGN and host galaxy classification. This yields a sample of $3\,154$ RLAGN in the Euclid Deep Field North, comprising $2\,803$ low-excitation radio galaxies (LERGs). We find that the LERG incidence as a function of stellar mass and radio luminosity follows a tight correlation over more than two orders of magnitude in stellar mass to $z = 3.0$. The slope of the power-law relation decreases from $\gamma \approx 2.5$ in the local Universe to $\gamma \approx 1.0$ at $z = 3.0$, indicating a relative increase in low-mass LERGs at earlier epochs. Separating the LERG population into actively star-forming and quiescent galaxies reveals that quiescent LERGs show a much steeper relation between incidence and stellar mass, with $\gamma\approx2.5$ compared to star-forming LERGs that show $\gamma\approx1.5$. However, the relation seems to be independent of whether galaxies are star-forming or quiescent at $z\geq2$, where both populations show relatively flat relations with $\gamma\approx1.0$. Our results suggest an evolutionary picture in which the LERG population transitions from cold-mode, merger-dominated accretion at earlier epochs to a scenario in which systems require a massive hot halo to fuel the AGN in the local Universe.
}


\titlerunning{Evolution of the radio-loud AGN fraction–stellar mass relation to $z\sim$3 in the EDF-N}
\authorrunning{S. Visser et al.}
   
\maketitle

\section{\label{sc:Intro}Introduction}
It is now well accepted that most massive galaxies host a \ac{smbh} at their centre \citep{Kormendy1995, Volonteri2012}, and that there is a tight correlation between the mass of the \ac{smbh} $M_\bullet$ and the bulge of galaxies in the local Universe. For example, $M_\bullet$ correlates with the bulge stellar velocity dispersion \citep{Ferrarese2000, Gebhardt2000ADispersion, Tremaine2002TheCorrelation}, and similarly, bulge mass or bulge luminosity shows a tight correlation with $M_\bullet$ (see \citealt{Kormendy2013} for a review). These correlations between \acp{smbh} and their host galaxies suggest co-evolution, often attributed to phases of \ac{agn} activity. 

During the \ac{agn} phase, the \ac{smbh} actively accretes material from its host galaxy and can release vast amounts of energy through radiation and matter outflows. Some \ac{agn} additionally launch relativistic, magnetically driven jets of hot plasma, which can deposit substantial amounts of kinetic energy into the surrounding medium \citep{Fabian2012}. \ac{agn} activity can couple its released energy to the ambient medium and regulate subsequent galaxy growth. This process, commonly referred to as \ac{agn} feedback, may suppress star formation in the host galaxy or cooling flows in galaxy groups and clusters, resulting in negative feedback. In some circumstances, however, \ac{agn} activity may instead enhance star formation, leading to positive feedback. These processes are therefore thought to play an important role in regulating galaxy evolution \citep{Fabian2012, Somerville2015PhysicalFramework, Harrison2024ObservationalInterpretation}.

Some \ac{agn} launch relativistic jets that interact with the ambient medium, producing non-thermal synchrotron emission as relativistic charged particles move through magnetic fields. This radio emission is unaffected by dust obscuration, making radio observations a particularly powerful and unique probe of jetted \ac{agn}, their environmental interactions, and their role in galaxy evolution. Wide-field radio observations such as the \acl{lofar} (\acs{lofar}; \citealt{VanHaarlem2013}) \acused{lofar} Two Meter Sky Survey (\acs{lotss}; \citealt{Shimwell2026TheRelease})\acused{lotss}, and the Karl G. Jansky \acl{vlass} (\acs{vlass}; \citealt{Perley2011THESCIENCE, Lacy2020TheDesign})\acused{vlass} enables the radio-source population to be characterised across unprecedented sky areas. The increased depth in deep radio surveys such as the \acs{vla}\acused{vla}-COSMOS \citep{Smolcic2017}, the \acl{mightee} (\acs{mightee}; \citealt{Jarvis2016TheSurvey, Hale2025MIGHTEE:1}), and the \ac{lofar} Deep Fields \citep{Tasse2020, Sabater2021, Bondi2024} probe much fainter radio populations, enabling studies of how \ac{rlagn} evolve with redshift \citep{Smolcic2017, Whittam2022, Kondapally2025}, while also probing \ac{sfg} and radio-quiet \ac{agn} \citep{Novak2017The5, Cochrane2023}.

Observations of \ac{rlagn} in the local Universe ($z\lesssim0.3$) have shown that they broadly separate into two accretion modes associated with different Eddington-ratio regimes: radiatively efficient (`radiative mode') and radiatively inefficient (`jet mode') \ac{agn} (e.g., \citealt{Heckman2014, Tadhunter2016, Hardcastle2020}). The latter are thought to host a radiatively inefficient \acl{adaf} (\acs{adaf}; \citealt{Narayan1994Advection-dominatedSolution})\acused{adaf}, with most of their energy released through two-sided collimated jets, whereas radiative-mode AGN emit predominantly electromagnetic radiation from a geometrically thin, optically thick accretion disc \citep{Shakura1973BlackSystems}. Radiative mode \ac{agn} are also observed to drive outflows, but the coupling of these outflows to the surrounding medium is poorly understood (e.g., \citealt{Harrison2024ObservationalInterpretation}).

Observationally, the jet and radiative mode populations are commonly classified as low- and high-excitation radio galaxies (\acs{lerg}s\acused{lerg} and \acs{herg}s\acused{herg}, respectively) based on their optical emission-line properties \citep{Heckman2014}. \acp{lerg} constitute the majority of the radio-loud AGN population and are typically hosted by massive, redder elliptical galaxies in dense environments, whereas \acp{herg} are less common and preferentially reside in bluer, less massive galaxies with more recent star formation. Note that this classification is not exclusive, such that some \acp{herg} are found in red, elliptical galaxies \citep[e.g.,][]{Janssen2012, Croston2019, Chilufya2025}.

The general interpretation is that \acp{herg} tend to accrete efficiently from cold gas by either external (e.g., galaxy mergers) or internal (e.g., bars or instabilities) processes, where a plentiful gas supply can lead to the formation of a standard accretion disc \citep[e.g.,][]{Allen2006TheGalaxies, Hardcastle2007HotGalaxies, Heckman2014}. For \acp{lerg}, the cooling of hot gas from the halo surrounding massive elliptical galaxies will lead to inefficient accretion \citep[e.g.,][]{Best2006AGN-controlledGalaxies, Heckman2014}. \acp{lerg} are often connected to negative \ac{agn} feedback, which keeps the star formation quenched and disrupts the circum-galactic medium when the jets expand from kpc up to Mpc scales \citep{Li2015COOLINGCLUSTERS, Gaspari2020LinkingFeedback, Oei2024BlackWeb}.

To understand the significance of radio-\ac{agn} feedback in galaxy evolution and \ac{agn} activity timescales, we need to study the redshift evolution of the \ac{rlagn} population. In the local Universe, one of the strongest observed relations is the dependence of the fraction of galaxies hosting an \ac{rlagn}, $f_\mathrm{RLAGN}$, on the host galaxy stellar mass, $M_*$. This relation has been studied extensively and is found to follow a power law \mbox{$f_\mathrm{RLAGN}\propto M_*^{2.5}$} for \acp{lerg} and \mbox{$f_\mathrm{RLAGN}\propto M_*^{1.5}$} for \acp{herg} \citep{Best2005a, Janssen2012, Sabater2019}. Interestingly, \cite{Sabater2019} found that $100\%$ of the most massive galaxies ($M_*>10^{11} M_\odot$) show radio-\ac{agn} activity. In other words, the nuclei of the most massive galaxies are always switched on. 

Other works extended this relation to $z=1.5$--$2.0$, using deep \ac{vla} and \ac{lofar} surveys, and found a flattening of the power-law relation with redshift \citep{Williams2015, Wang2024Cosmic4, Kondapally2025}. In addition, some works have found that the \ac{lerg} incidence in \ac{sfg} at higher redshifts is more comparable to that of \acp{herg} in the local Universe with \mbox{$f_\mathrm{RLAGN}\propto M_*^{1.5}$}. Alternatively, quiescent \acp{lerg} show \mbox{$f_\mathrm{RLAGN}\propto M_*^{2.5}$}, which can explain the flattening of the \ac{rlagn} incidence with redshift \citep{Kondapally2022, Kondapally2025}. These results indicate that the standard paradigm of classifying \acp{herg} and \acp{lerg} based on their accretion rate may not always hold, suggesting a connection between \acp{lerg} and the accretion of cold gas \citep{Kondapally2025, Q1-SP065}. 

However, reliable classification of \acp{herg} and \acp{lerg} suffers from selection bias. \cite{Arnaudova2025} compared spectroscopic classifications to state-of-the-art photometric classifications from \ac{sed} fitting, and found that these show reasonable overall agreement. Nevertheless, they found significant differences, including identifying $2$--$5$ times more radio-quiet \ac{agn} using spectroscopic data. This may impact the separation between \acp{herg} and \acp{lerg}, causing their accretion properties to appear less distinct \citep{Whittam2022}. Recent results on the X-ray properties of \acp{herg} and \acp{lerg} have, however, shown that \acp{lerg} tend to have much lower average X-ray luminosities (a proxy for the accretion rate) compared to the \acp{herg} \citep{Kondapally2026}. Because studies in the local Universe have found a significantly shallower slope for \acp{herg} \citep{Janssen2012}, it is crucial to exclude \acp{herg} contamination before claiming a flattening of the \mbox{$f_\mathrm{RLAGN}$--$M_*$} relation for \acp{lerg}.

While large radio surveys offer tremendous scientific potential, understanding the physical processes driving radio-source activity requires detailed characterisation of their host galaxies through multi-wavelength observations. Combining wide-field radio surveys with new facilities such as \Euclid \citep{EuclidSkyOverview} will revolutionise our ability to survey wide areas with high sensitivity. These facilities will deliver increased source statistics and detailed host-galaxy measurements to unprecedented depths. Such data enable large statistical studies of radio-source populations and their connection to host-galaxy properties, environments, and redshift. 

In this work, we investigate how the incidence of \acp{lerg} evolves with $M_*$, $z$, and radio luminosity. We focus on the \ac{edfn} by combining deep \ac{lofar} radio data \citep{Bondi2024} with the new optical/\ac{nir} \Euclid data \citep{Q1-TP001}. We obtain the sensitivity and host-galaxy characterisation required to probe to $z=3.0$ and mass-completeness down to $M_*\approx10^{11} M_\odot$. A key advantage of these data is the substantial improvement in $M_*$ completeness provided by \Euclid\ compared to previous analyses of other \ac{lofar} deep fields. This enhanced depth is crucial for investigating the \ac{lerg} incidence at lower $M_*$ and higher $z$. Consequently, we can extend the \mbox{$f_\mathrm{RLAGN}$--$M_*$} relation beyond the epoch of cosmic noon, at which both the \ac{sfr} density and the \ac{smbh} growth rate in the Universe peaked \citep{Shankar2009Self-consistentEfficiency, Madau2014}.

The paper is structured as follows. Section~\ref{sc:Data} describes the \ac{lofar} and \Euclid data in the \ac{edfn}. Section~\ref{sc:Cross-matching} presents the methods used to cross-match the \ac{lofar} and \Euclid source catalogues, the $M_*$ completeness for our sample, and how to mitigate potential contamination. This is followed by the selection of \ac{rlagn}, the differentiation between \acp{lerg} and \acp{herg}, and the method to calculate \ac{lerg} fractions in Sect.~\ref{sc:RLAGN selection}. In Sect.~\ref{sc:RLAGN fractions}, we present the \ac{lerg} incidence as a function of $M_*$ and $z$ for different radio luminosity limits, and discuss the reliability and interpretation of our results. Finally, Sect.~\ref{sc:Conclusions} presents our conclusions. In this work, we use a flat $\Lambda$CDM cosmology with $H_0=70\,\kmsMpc$ and $\Omega_\mathrm{m}=0.3$, and assume a radio spectral index of $\alpha=-0.7$, with \( S_\nu \propto \nu^{\,\alpha} \).

\section{\label{sc:Data} Data}
We first describe the \ac{lofar} radio data in the \ac{edfn}. We then present the \Euclid data and catalogues, followed by the supplementary data used to obtain physical parameters and spectroscopic redshifts where available.

\subsection{\label{sub-sc:LOFAR data} LOFAR data in the EDF-N}
The \ac{edfn} is the fourth \ac{lofar} Deep Field and has a sensitivity similar to that of the other three fields -- ELAIS-N1, Lockman Hole, and Bo\"{o}tes -- presented in \ac{lotss} Deep Fields DR1 \citep{Tasse2020, Sabater2021, Best2023}. The \ac{edfn} was observed with the \ac{hba} at 144 MHz for $400$ hours. When fully processed at \ang{;;6} resolution, the \ac{edfn} will have a central \ac{rms} noise of $10 \,\si{\uJy}\,\mathrm{beam}^{-1}$. An initial image and a corresponding radio source catalogue in the \ac{edfn} were obtained by \cite{Bondi2024} from $72\, \rm h$ of observations. This image has an angular resolution of \ang{;;6} and a central \ac{rms} noise of $32 \, \si{\uJy}\,\mathrm{beam}^{-1}$.

\cite{Bondi2024} extracted the radio source catalogue using the \acl{pybdsf} (\acs{pybdsf}, \citealt{Mohan2015})\acused{pybdsf}, which is a tool to decompose radio interferometry images into a set of islands above a given \ac{snr}. These islands are fitted using Gaussians, shapelets, or wavelets, giving not only information about the measured fluxes but also about the morphology of radio sources. The resulting catalogue contains sources with a peak flux satisfying $\mathrm{S/N} > 5$ in a region of approximately \mbox{$10 \,\mathrm{deg}^2$} centred at \mbox{$\mathrm{RA}=\ra{17;58;55.2}$}, \mbox{$\mathrm{Dec}=\ang{+66;01;12.0}$}.

In this paper, we make use of the updated radio source positions in the catalogue for the \ac{edfn} as presented by \cite{Bisigello2025}, hereafter \citetalias{Bisigello2025}\footnote{\url{https://lofar-surveys.org/deepfields_public_edfn.html}}. To enhance their cross-matching procedure, \citetalias{Bisigello2025} improved the positional accuracy of radio sources by using the high-resolution image of the \ac{edfn} made by the International \ac{lofar} Telescope. This image has been produced following the methods described in \cite{Sweijen2022} and \cite{deJong2024} using $32\, \rm h$ of the $72\, \rm h$ available. The final high-resolution images cover the inner $6.25 \, \mathrm{deg}^2$ region of the \ac{edfn} reaching a central sensitivity of $40 \, \si{\uJy}\,\mathrm{beam}^{-1}$ and an angular resolution of \ang{;;1.5}. The final updated source catalogue comprises $23\,309$ radio sources \citetalias{Bisigello2025} and will be used to cross-match with the \Euclid data.

\subsection{\label{sub-sc:Euclid data} \Euclid data}
Complementing the radio data, we use the first Quick Release of \Euclid data (Q1; \citealt{Q1cite}). This release comprises the \acl{edf}\acused{edf} (EDF-North, EDF-South, and EDF-Fornax) covering a total area of $63.1 \,\mathrm{deg}^2$. All \ac{edf} have been observed in the four \Euclid photometric bands, that is, one optical (\IE) using the VIS instrument and three \ac{nir} bands (\YE, \JE, and \HE) using the \acl{nisp} \citep[\acs{nisp}\acused{nisp};][]{EuclidSkyNISP}. \Euclid provides high angular resolution images with a mean \ac{fwhm} of \mbox{$\ang{;;0.158}\pm\ang{;;0.001}$} captured by the VIS instrument \citep{EuclidSkyVIS} for all galaxies down to $\IE = 25.5$ \citep{Q1-TP002}. This allows for detailed characterisation of galaxy morphologies to large distances (e.g., \citealt{Q1-SP013, Q1-SP040, Q1-SP047}).

The observations from the VIS and \ac{nisp} instruments are complemented by ground-based photometry (EXT data) to form the multi-wavelength MER catalogue \citep{Q1-TP004}. Sources are detected using \texttt{SourceExtractor++} \citep{Bertin2020TheSoftware} independently for the VIS and an \ac{nir}-stack detection mosaic. Thus, the MER catalogue includes only sources detected by \Euclid. The \texttt{VIS{\_}DET} column indicates VIS detection: \texttt{VIS{\_}DET}=1; otherwise, the source is detected only in the stacked \ac{nir} image (\texttt{VIS{\_}DET}=0). 

For this work, we focus on the data in the \ac{edfn}, since this overlaps with the \ac{lofar} deep field. The \ac{edfn} observed by \Euclid is centred at $\mathrm{RA}=\ra{17;58;55.9}$, $\mathrm{Dec}=\ang{+66;01;03.7}$ and covers an area of $22.9 \, \deg^2$, which is over two times larger than the \ac{lofar} pointing in the \ac{edfn}. To obtain \Euclid counterparts for the \ac{lofar} source catalogue described in, we retrieved $8\,868\,897$ \Euclid sources in the \ac{edfn} from the MER catalogue satisfying the following criteria\footnote{For a full description of the \Euclid Data Products: \url{https://euclid.esac.esa.int/dr/q1/dpdd/}}:
\begin{itemize}
    \item \texttt{SPURIOUS{\_}FLAG} = 0, to filter out spurious sources;
    \item \texttt{PHZ{\_}CLASSIFICATION} $\neq 1$, to exclude objects classified as a star by the PHZ processing function based on their photometry \citep{Q1-TP005};
    \item \texttt{DET{\_}QUALITY{\_}FLAG} = $0 \,|\, 2 \,|\, 512$, to select sources with good quality photometry, blended sources that have been successfully deblended, or sources within the area of an extended object \citep{Q1-TP004};
    \item $\ac{snr}\geq3.0$ for $\mathrm{\texttt{FLUX{\_}DETECTION{\_}TOTAL}}$, to ensure a $3\,\sigma$ detection in \IE ($\texttt{VIS{\_}DET}=1$) or in the stacked \ac{nir}-stacked image ($\texttt{VIS{\_}DET}=0$).
\end{itemize}

Accurately investigating the redshift evolution of \ac{rlagn} requires reliable redshift estimates. In addition to the spectroscopic redshifts provided by \Euclid\ \citep{Q1-TP007}, Q1 benefits from extensive overlap with existing spectroscopic surveys. To maximise the spectroscopic completeness of our sample, we use the catalogue described by \cite{Q1-TP005}, which matches \Euclid\ MER sources to approximately 40 external spectroscopic surveys. Specifically, we employ an updated version of this catalogue (Euclid Collaboration: Saulder et al. in prep.), comprising $47\,466$ highly reliable spectroscopic redshifts in the \ac{edfn} with $0.0 \leq z \leq 5.8$, with the majority originating from the \ac{desi} DR1 survey \citep{Collaboration2025h}.

For host-galaxy photometric redshifts and physical properties, we adopt the parameters derived by \cite{Q1-SP031}, hereafter \citetalias{Q1-SP031}. They used the \Euclid pipeline's \ac{nn} algorithm, adding two further bands from \acl{irac} (\acs{irac}, \citealt{Fazio2004Telescope})\acused{irac} onboard \textit{Spitzer} \citep{Werner2004Mission} and applying more constrained priors to improve the quality of their \ac{sed} fitting results. For sources with spectroscopic redshifts from \ac{desi}, \citetalias{Q1-SP031} compared their estimates with those in the \ac{desi} value-added catalogue \citep{Siudek2024Value-addedSurvey}. They found that the photometric redshifts achieve a \ac{nmad} of $\sigma_z/(1 + z) \approx 0.04$ up to $z = 3$, while $M_*$ and \ac{sfr} estimates have an \ac{nmad} of $\approx 0.35$ and $\approx 0.55$ dex, respectively.

The reliability of photometric redshifts and physical parameters derived from \ac{sed} fitting can be significantly affected by the presence of an \ac{agn}. In such systems, an a priori unknown fraction of the observed flux originates from the \ac{agn} rather than the stellar and interstellar components of the host galaxy, potentially biasing estimates of galaxy properties. To mitigate these effects, \cite{Q1-Laloux}, hereafter \citetalias{Q1-Laloux}, derived \ac{agn}-specific physical parameters. 

First, \citetalias{Q1-Laloux} identified \ac{agn} candidates using the range of selection techniques developed for Q1 \citep{Q1-SP027}. The next step was to obtain reliable redshifts for all \ac{agn} candidates. Whenever spectroscopic measurements were unavailable, improved redshift estimates were obtained through visual inspection of \Euclid \ac{nisp} slitless spectroscopy \citep{Q1-SP068}, machine-learning photometric redshifts for point-like X-ray selected sources \citep{Q1-SP003}, and template-fitting photometric redshifts derived with the Bayesian code \texttt{Phosphoros}. For the latter, the luminosity prior and template library were modified to improve redshift estimates for AGN and reduce the rate of catastrophic outliers. Lastly, the resulting physical parameters were obtained through a dedicated \ac{sed}-fitting campaign using \acs{cigale} (\acl{cigale}; \citealt{Boquien2019CIGALE:Emission, Yang2022FittingCode})\acused{cigale}. In particular, \citetalias{Q1-Laloux} employed the \texttt{SKIRTOR} \ac{agn} module \citep{Stalevski2011Three-dimensionalMedium, Stalevski2016TheNuclei}, which models a two-phase dusty torus consisting of dense clumps embedded within a lower-density smooth component. In this work, we adopt the physical parameters from this catalogue for \ac{agn} candidates, while using the measurements from \citetalias{Q1-SP031} for all other sources, as described in Sect.~\ref{sc:RLAGN selection}.

\section{\label{sc:Cross-matching} Cross-matching LOFAR and \Euclid source catalogues}
In this section, we discuss the \ac{lr} method that was used to identify \Euclid counterparts for the \ac{lofar} sources in the \ac{edfn}. The following subsections then discuss the $M_*$ completeness limit for \Euclid galaxies and an estimation of the contamination in our cross-matched sample as a function of $M_*$.

\subsection{\label{sub-sc: matching_method} Likelihood ratio identifications}
As will be discussed in more detail in Sect.~\ref{sub-sc: matching contamination}, simply cross-matching the \ac{lofar} and \Euclid source catalogues using an \ac{nn} approach will lead to significant contamination of the cross-matched sample. This effect becomes more pronounced towards smaller $M_*$ because the galaxy number density increases for lower masses. Since the \ac{rlagn} incidence can reach low values of $f_\mathrm{RLAGN}\sim 10^{-4}$ for $M_* < 10^{10} \, M_\odot$ \citep[see e.g.,][]{Janssen2012, Williams2015, Sabater2019}, it is crucial to both mitigate and account for this contamination.

As mentioned in Sect.~\ref{sub-sc:LOFAR data}, \citetalias{Bisigello2025} used the \ac{lr} method, incorporating colour and magnitude, to identify optical/\ac{nir} counterparts to the \ac{lofar} \ac{edfn} catalogue. The \ac{lr} method is a statistical method to identify counterparts in images with widely differing frequency ranges \citep{DeRuiter1977, Sutherland1992}. The \ac{lr} is the ratio of the probabilities that a galaxy is the true counterpart or an unrelated background object, given its properties. As the Euclid Q1 data were unavailable, \citetalias{Bisigello2025} could not use them for cross-matching. Here, we apply the \ac{lr} method for the \Euclid data to explore its capabilities in future data releases. We follow the method in appendix B of \cite{Best2003}, which makes use of the coordinates and the positional uncertainties of both radio and optical sources (for more information, see \citealt{Richter1975, DeRuiter1977}).

We avoid using galaxy colour information, as in \citetalias{Bisigello2025}, to avoid introducing priors on galaxy colour or magnitude that could favour particular host-galaxy populations. Although previous studies have shown that \ac{lofar} sources tend to have redder counterparts than the general galaxy population in the local Universe \citep{Williams2019}, recent studies of the \ac{lofar} Deep Fields have also identified radio sources associated with blue galaxies \citep{Best2023, Igo2024TheConnection}. Including colour or magnitude information can improve counterpart identification, but may also favour populations represented by the adopted priors. This is particularly relevant given the correlation between galaxy colour and stellar mass \citep{Baldry2006, Grutzbauch2011}, which could bias analyses of host-galaxy properties. A positional-only \ac{lr} approach avoids these additional photometric priors, and we test its validity in Sect.~\ref{sub-sc: LR biases}.

The \ac{lr} method uses an error-weighted distance between an optical and a radio source defined as 
\begin{equation} \label{eq:r_def}
    r\equiv\left(\frac{\Delta \alpha^2}{\sigma_\alpha^2}+\frac{\Delta \delta^2}{\sigma_\delta^2}\right)^{1 / 2} \; ,
\end{equation}
where $\Delta \alpha$ and $\Delta \delta$ are the differences in right ascension $\alpha$ and declination $\delta$ between the radio and optical coordinates, and $\sigma_\alpha$ and $\sigma_\delta$ are the associated errors. These errors are the quadratic sum of the positional errors for the radio source, the optical source and the astrometric uncertainty or offset between both surveys, \mbox{$\sigma^2=\sigma^2_\mathrm{ast}+\sigma^2_\mathrm{opt}+\sigma^2_\mathrm{rad}$}. The lower radio resolution makes the radio positional uncertainties dominant. Each radio source has unique positional uncertainties calculated by \ac{pybdsf}, which are reported in the radio source catalogue. For the radio source coordinates, we adopt the corrected coordinates obtained by \citetalias{Bisigello2025} after an astrometric correction. For the \Euclid coordinates, we estimate the positional uncertainty as \mbox{$\mathrm{FWHM}\,/\, \ac{snr} = \ang{;;0.18}\,/\, \ac{snr}$} as in \cite{Ivison2007}. We conservatively adopt the 99th percentile, yielding \mbox{$\sigma_{\alpha,\,\mathrm{opt}} = \sigma_{\delta,\,\mathrm{opt}}= \ang{;;0.06}$}. Lastly, for the astrometric uncertainty between the two surveys, we use \mbox{$\sigma_{\alpha,\,\mathrm{ast}} = \sigma_{\delta,\,\mathrm{ast}} = \ang{;;0.47}$}, which is the median astrometric radio-optical offset for the \ac{edfn} found by \citetalias{Bisigello2025}. These values account for any remaining astrometric offset for individual sources and were also used by \citetalias{Bisigello2025}, who found them representative of astrometric variations across the \ac{edfn}, facilitating comparison with our results.

Assuming a random distribution of radio sources across the \ac{edfn}, the probability of finding the nearest unrelated object at a separation between $r$ and $r+ \diff r$ follows a Poissonian background distribution determined by the source surface density $\rho$. This assumption implicitly neglects spatial variations in the source density and angular clustering, which may be significant in overdense environments such as galaxy clusters. A uniform background density may therefore underestimate chance associations in these regions. Nevertheless, the Poissonian approximation provides a useful baseline for estimating chance associations. In contrast, the distribution of true counterparts is expected to follow a Rayleigh profile set by the positional uncertainties. The \ac{lr} for each radio–optical association is defined as the ratio of the probability of a true counterpart to that of a random association at a distance between $r$ and $r+ \diff r$. For verification purposes, we compared the observed distribution of radial offsets between \Euclid\ and LOFAR sources and find that it is well described by a Rayleigh distribution at separations below \mbox{\ang{;;1}}, while larger separations are increasingly dominated by chance associations.

Before computing the \ac{lr}, we account for bright stars that can produce diffraction spikes in optical and \ac{nir} images, leading to unreliable photometry. For this purpose, we use the star mask constructed by \citetalias{Bisigello2025} for the radio source catalogue in the \ac{edfn}. The masking procedure follows the approach outlined in section~3.4.3 of \cite{Kondapally2021}, which used the \textit{Gaia} DR2 catalogue to define magnitude-dependent masking radii for individual stars. This results in circular masks with radii ranging from \ang{;;178} for stars with $g \leq 8.50$ to \ang{;;2} for stars with $g_\mathrm{HSC} \geq 21.1$, yielding a total masked area of $1.856 \,\rm deg^2$ in the \ac{lofar} \ac{edfn}. After applying the mask, the \ac{lofar} \ac{edfn} field covers an area of $8.73\,\mathrm{deg}^2$ with $19\,550$ \ac{lofar} sources.

The \ac{lr} was calculated for every single-component \ac{lofar} source \mbox{(\texttt{S$\_$code = S})} with a major axis smaller than \ang{;;15} following \cite{Kondapally2021} and \citetalias{Bisigello2025}. We excluded sources for which \ac{pybdsf} was not able to extract an accurate size estimate. This occurs when radio sources are blended, or when visual inspection shows that two lobes classified as separate sources by \ac{pybdsf} belong to the same source \citep{Bondi2024}. These selections retain $17\,971$ out of $19\,550$ ($91.9\%$) radio sources suitable for \ac{lr} identification. The matching of the excluded sources will be discussed later in this section. Regarding the \Euclid sources, we computed the \ac{lr} for all galaxies satisfying $r\leq3.0$ (see Eq.~\ref{eq:r_def}). Assuming that sources are randomly distributed across the field, $r\leq3.0$ means that there is a $3.3\%$ chance that the real optical/\ac{nir} counterpart is further away.

To find an appropriate threshold value $L$ to select reliable matches with $\mathrm{LR}\geq L$, we estimate the completeness $C(L)$, the fraction of real identifications that are accepted, and the reliability $R(L)$, the fraction of accepted identifications that are correct \citep{DeRuiter1977, Best2003}
\begin{equation}
    C(L)=1-\frac{1}{N_\mathrm{id}} \sum_{\mathrm{LR}_i<L} p_i(\mathrm{id} \,|\, r) \; ; \; 
    R(L)=1-\frac{1}{N_\mathrm{id}} \sum_{\mathrm{LR}_i \geq L} p_i(\mathrm{b} \,|\, r)\; .
\end{equation}
Here, $p_i(\mathrm{b} \,|\, r)$ and $p_i(\mathrm{id} \,|\, r)$ are the probabilities that source \(i\) is an unrelated background object `b' or a real counterpart `id', respectively, and $N_\mathrm{id}$ is the total number of real identifications obtained by summing $p_i(\mathrm{id} \,|\, r)$ for all sources. For an object $i$, we need to calculate these posterior probabilities using
\begin{equation} \label{eq:LR probabilities}
    p_i(\mathrm{id} \,|\, r)=\frac{\phi \,\mathrm{LR}_i}{\phi \,\mathrm{LR}_i+1-\phi}\; , \;
    p_i(\mathrm{b} \,|\, r)= 1 - p_i(\mathrm{id} \,|\, r)\; ,
\end{equation}
where $\phi$ is the a priori probability of finding an optical counterpart for a radio source. To estimate $\phi$, we calculated the distance between all single-component \ac{lofar} sources and their nearest \Euclid galaxy. As discussed above, this showed that the distribution is dominated by a Rayleigh distribution at short distances and the majority ($15\,960$ sources) of the matches are found within a \ang{;;1.5} radius. We use this number to estimate the a priori probability to be $\phi=15\,960/17\,971=88.8\%$. 

To study the \ac{rlagn} incidence down to $M_* < 10^{10} \, M_\odot$, where the \ac{rlagn} fraction is relatively small, we require stringent cross-matching with minimal contamination. We found that the contamination varies little with $L$. This is caused by the stringent requirement of $r<3.0$. We adopt the threshold value $L$ for which \mbox{$C(L)=R(L)$}, resulting in $L=0.462$ with both the completeness and the reliability being $96.7\%$. Using this threshold value, we find $13\,661$ matches with the \ac{lr} method.

\subsection{Testing for biases in LR identifcations \label{sub-sc: LR biases}}
In an \ac{lr} framework that does include magnitude priors and colour information, counterparts at very small separations naturally obtain high \ac{lr} values, such that the colour term contributes little to the final association probability. Colour information mainly becomes important when multiple candidate counterparts are present at comparable separations, where it preferentially selects sources with colours commonly associated with radio counterparts \citep[e.g.,][]{Williams2019, Kondapally2021}. To evaluate the impact of neglecting colour information, we selected all radio sources with two optical candidate counterparts satisfying $r \leq 3.0$, yielding $1\,241$ objects ($6.8\%$ of the sources suitable for the \ac{lr} analysis). Of these, $896$ have detections in both \IE\ and \HE, allowing us to construct the colour $\IE - \HE$. We then divided the sample into two categories: a `clear-match' class, containing sources with one close candidate ($r \leq 1.0$) and a second candidate at larger separation ($2.0 \leq r \leq 3.0$), and an `ambiguous' class, where both candidates satisfy $2.0 \leq r \leq 3.0$. These classes contain $191$ ($1.1\%$) and $148$ ($0.8\%$) radio sources, respectively. The colour distribution of the counterparts showed only marginal differences between the two classes. More importantly, these ambiguous systems constitute only a very small fraction of the total sample, indicating that neglecting colour priors does not significantly affect the matching results or the rest of our analysis. We also note that only $10$ ($0.06\%$) radio sources have two candidate optical counterparts with $r \leq 1.0$, such that we do not expect this to affect our classification significantly.

For the $1\,579$ radio sources ($8.1\%$ of the full sample) that are non-compact ($7.5\%$), too large ($1.2\%$), or lacking a reliable size estimate ($0.7\%$), we adopt the optical identifications from \citetalias{Bisigello2025}. They visually inspected sources that could not be identified using the \ac{lr} method (see section 3.5 of \citetalias{Bisigello2025}). We then cross-matched their optical/IR counterpart coordinates with the \Euclid catalogue using an \ac{nn} search radius of \mbox{\ang{;;0.7}}.\footnote{The expected number of spurious sources for this radius is \mbox{$1579\,\pi\,(1.9\times10^{-4} \,\mathrm{deg})^2\,(3.7\times10^5 \,\mathrm{deg}^{-2}) \approx 69$}, corresponding to an estimated misidentification chance of \mbox{$69/1579\approx4.35\%$}.} This radius was motivated by the separation distribution between \Euclid sources and the counterparts identified by \citetalias{Bisigello2025}, which peaks below \ang{;;0.7}. At larger separations, match counts increased as expected for random associations because of the larger search area. The \ac{nn} cross-match adds $1\,317$ sources to our sample with a median separation of \mbox{\ang{;;0.151}}, meaning that they are highly reliable. Ultimately, this results in a fully cross-matched sample in the \ac{edfn} of $14\,978$ sources, yielding $14\,978/19\,550=76.6\%$ of the \ac{lofar} sample with a counterpart.

Using the \ac{lr} method with positional information alone could yield more unreliable identifications. Namely, colour priors can help find a reliable counterpart when the distance criterion is ambiguous. However, we prefer to avoid the introduction of potential biases, e.g. through the correlation between galaxy colour and $M_*$. To verify the cross-matching, we visually inspected a random subset of 100 \ac{lr}- and 100 \ac{nn}-matched radio sources.\footnote{These represent approximately $1\%$ and $10\%$ of the total number of \ac{lr}- and \ac{nn}-matched sources, respectively.} Approximately $90\%$ of the \ac{lr} matches were correctly identified, while all \ac{nn} matches were confirmed, as expected since they had already been visually inspected by \citetalias{Bisigello2025}. Comparing our \ac{lr} matches with those of \citetalias{Bisigello2025} yielded an agreement of $92\%$. In nearly all discrepant cases, two \Euclid\ sources were separated by less than the \ang{;;6} resolution of the radio maps, making it difficult to unambiguously identify the correct counterpart.

\subsection{\label{sub-sc:Final sample} Final sample with physical parameters}
The last step in creating a cross-matched \ac{lofar}-\Euclid source catalogue is to obtain the \ac{sfr} and $M_*$ for the $14\,978$ galaxies from the catalogue in \citetalias{Q1-SP031} or in \citetalias{Q1-Laloux} if they are \ac{agn} candidates. The classification of sources into \ac{agn} and non-\ac{agn} is discussed in more detail in Sect.~\ref{sc:RLAGN selection}. Since \citetalias{Q1-SP031} applied additional selections to the Q1 data, and not all sources were detected by \ac{irac}, our full matched sample contains $13\,502$ sources with physical parameter estimates. Subsequently, we require reliable physical parameters with \texttt{nnpz{\_}flags}$ = 0$ for non-\ac{agn} or \mbox{$\chi_\mathrm{red}^2 \leq 5.0$} for the \ac{sed} fits of the \ac{agn} candidates, which reduces the total number of sources to $12\,032$. Lastly, we impose a cut to retrieve only those sources in the redshift range $0.03 \leq z \leq 3.0$ and satisfying $\mathrm{SFR} \geq 10^{-5} \, M_\odot \, \mathrm{yr}^{-1}$ to ensure the reliability of the photometric redshifts and derived physical properties (\citetalias{Q1-SP031}). 

These selections result in a total of $10\,794$ sources in our final sample, comprising $6\,161$ sources with parameters from \citetalias{Q1-SP031} and $4\,633$ sources with parameters from \citetalias{Q1-Laloux}. $3\,653$ sources have physical parameters available in both catalogues, enabling a direct comparison of the derived quantities (see Sect.~\ref{sub-sc:SF vs quiescent} and Appendix~\ref{app: quiescent and sf lerg fractions}). We apply the same selection criteria to the full sample of $3\,877\,025$ \Euclid galaxies in the overlapping area between \ac{lofar} and \Euclid in the \ac{edfn}. This resulted in a total of $1\,945\,312$ sources, of which $1\,838\,361$ were drawn from \citetalias{Q1-SP031} and $106\,951$ from \citetalias{Q1-Laloux}. Note that $3\,231$ ($29.9\%$) sources in our cross-matched sample and $34\,665$ ($1.8\%$) sources in the galaxy sample have a spectroscopic redshift.

Using the redshift for a source with flux density $S_{144}$ and applying the standard radio $K$-correction, we calculate the rest-frame radio luminosity at $144$ MHz
\begin{equation}
L_{144} = 4\,\pi\, \DL^2\,S_{144}\,\left(1+z\right)^{-\alpha-1} \,\rm W \, Hz^{-1}\; ,
\end{equation}
where $\DL$ is the luminosity distance and $\alpha=-0.7$ is the typical radio spectral index with a dispersion of $0.2$--$0.3$ \citep[e.g.,][]{Bohme2023}. Based on early \ac{lofar} \ac{hba} observations in the \ac{edfn}, \cite{White2024} found this spectral index to be representative of the radio source population.

\subsection{\label{sub-sc: matching contamination} Estimating the cross-matching contamination fraction}
As a result of \Euclid's excellent sensitivity, the \ac{edfn} has an average galaxy number density of \mbox{$3.7 \times 10^5 \,\rm{deg}^{-2}$}, mostly consisting of low-mass objects. Therefore, a simple \ac{nn} cross-match between the (lower-resolution) \ac{lofar} and Q1 sources results in many low-mass objects being identified as counterparts, thus yielding relatively high misidentification rates. The \ac{lr} method described in Sect.~\ref{sub-sc: matching_method} reduces these rates, but we further estimate the contamination fraction of the cross-matched sample as a function of $M_*$ to account for residual effects.

\begin{figure}[htbp!]
    \centering
    \includegraphics[width=0.49\textwidth]{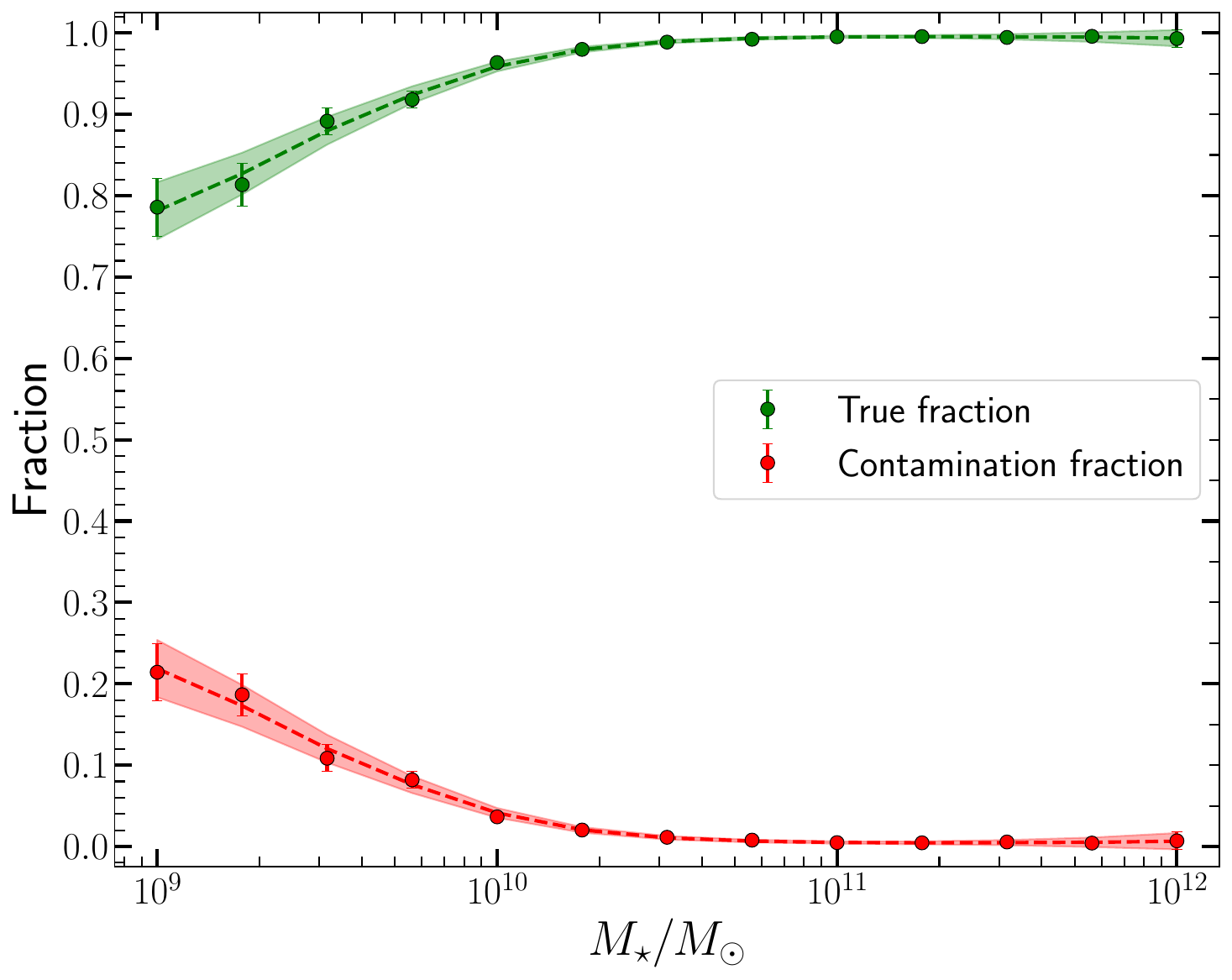}
    \caption{True (green) and contamination (red) fractions for the \ac{lofar}--\Euclid cross-matching as a function of $M_*$ obtained with the \ac{lr} method. For each, we show the interpolated curve (dashed line) and the $68\%$ confidence interval (shaded region).}
    \label{fig:contamination fraction}
\end{figure}

To obtain the contamination fraction, i.e., the chance of matching with a source that is not the genuine counterpart, as a function of $M_*$, we again perform the \ac{lr} matching after separating the \Euclid sources into $M_*$ bins using the same \ac{lr} threshold and maximum value for $r$ as used in Sect.~\ref{sub-sc: matching_method}. Next, we calculate the mean of $p_i(\mathrm{id} \,|\, r)$ and $p_i(\mathrm{b} \,|\, r)$ for every $M_*$ bin, which we define as the true fraction $f_{\rm true}(M_*)$ and the contamination fraction $f_{\rm contam}(M_*)$, respectively. The results of this analysis can be seen in Fig.~\ref{fig:contamination fraction} and show that especially for $M_* \leq 10^{10} \,M_\odot$ there is a significant contamination fraction rising from about $5\%$ at $M_* \approx 10^{10} \,M_\odot$ to $>20\%$ at $M_* \leq 10^{9} \,M_\odot$.

To verify the true and contamination fractions obtained from the \ac{lr} matching, we also performed a Monte Carlo-based analysis. For all optical sources in a given $M_*$ bin, we calculate the matching fraction for randomly sampled radio coordinates in the \ac{edfn} using a radius \mbox{$(\sigma_\mathrm{\alpha, \, rad}^2 + \sigma_\mathrm{\delta,\, rad}^2)^{1/2}$} to take into account that sources with a higher \ac{snr} have a more well-defined position. We then calculate the matched fraction for the real radio coordinates, such that the difference between the real and random matched fractions yields the contamination fraction. Within uncertainties, the results of this analysis are the same as the results seen in Fig.~\ref{fig:contamination fraction} obtained from the \ac{lr} matching.

\subsection{\label{sub-sc: Mstar completeness} Stellar mass completeness for Q1}
\begin{figure}[htbp!]
    \centering
    \includegraphics[width=0.49\textwidth]{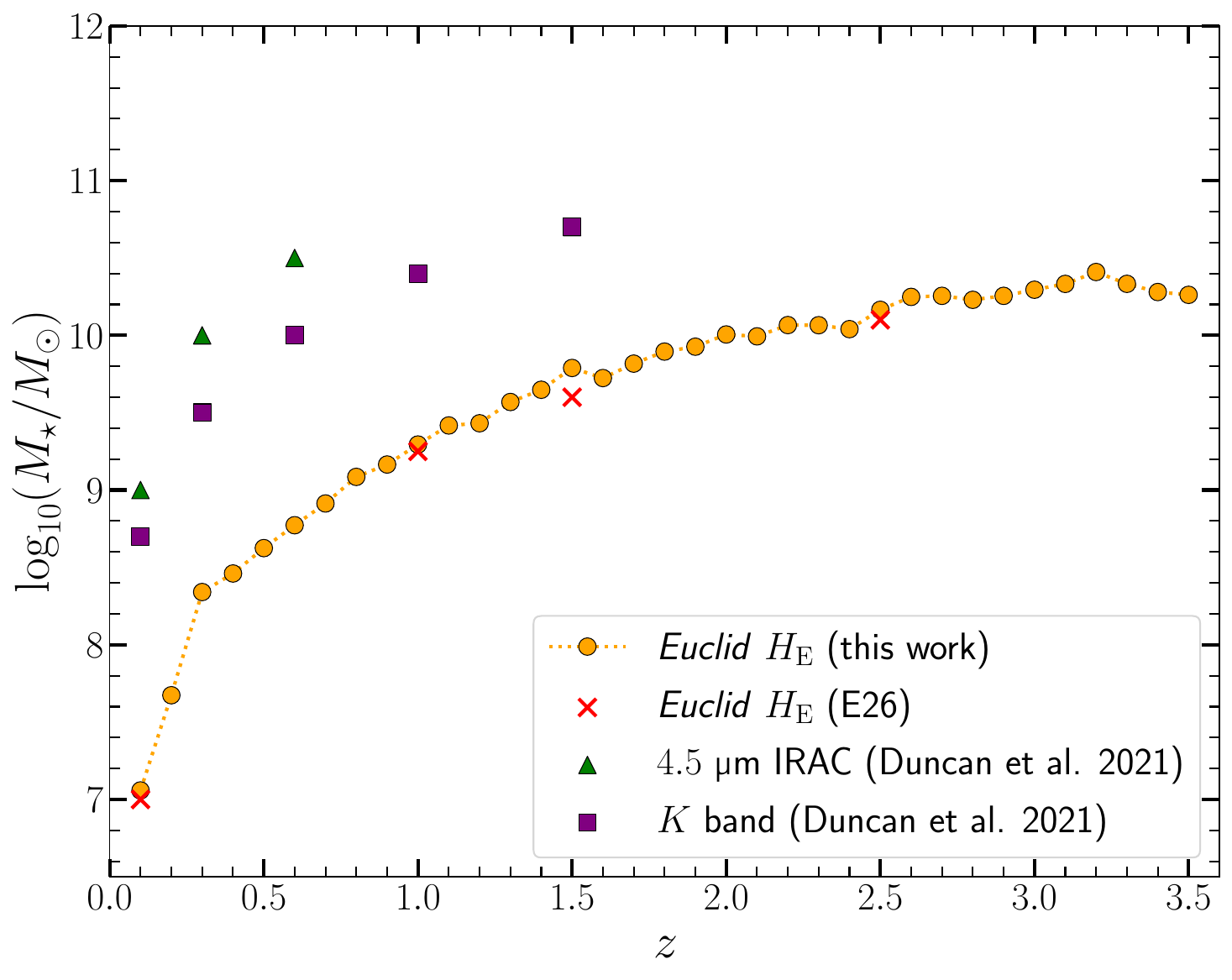}
    \caption{Stellar mass completeness limit as a function of $z$ for \Euclid galaxies in the \ac{edfn}. The orange line shows the $90\%$ completeness limit using galaxies with $\HE < 24$. Red crosses show the $95\%$ completeness limit reported in \citetalias{Q1-SP031}. Purple squares and green triangles show the $90\%$ completeness limits obtained for the ELAIS-N1 field using the $K$ and the $4.5 \, \micron$ \ac{irac} band, respectively \citep{Duncan2021}.}
    \label{fig:mass_completeness}
\end{figure}

To prevent any bias on the \ac{rlagn} incidence as a function of $M_*$, we need to quantify the minimum $M_*$ above which Q1 is complete, given its limiting magnitude. We use the method described in \cite{Pozzetti2010} and select the faintest $20\%$ of the \Euclid galaxy sample in the \ac{edfn} limited to $\HE < 24$, which is the $5\,\sigma$ limit for \HE-detected sources. To verify this limit, we estimated the $90\%$ magnitude completeness limit by fitting a powerlaw distribution to the observed number counts for \mbox{$\ac{snr}\geq5$} \HE-detected sources. The $90\%$ magnitude completeness limit, defined as the magnitude at which the observed number counts drop to $90\%$ of the fitted power-law distribution, is $\HE = 23.97$, in agreement with the $5\,\sigma$ detection limit. 

Subsequently, we calculated the stellar mass each galaxy would have if its apparent magnitude were equal to the limiting magnitude given its redshift, $M_\mathrm{lim}$, using \citep{Pozzetti2010}
\begin{equation}
    \log_{10}(M_\mathrm{lim}/M_\odot) = \log_{10}(M_*/M_\odot)\, +\, 0.4\,(\HE - 24)\; .
\end{equation}
To obtain the $M_*$ completeness limit as a function of redshift, $M_\mathrm{lim}(z)$, we divided the sample into redshift bins and determined the 90th percentile of $M_\mathrm{lim}$ in each bin. Figure~\ref{fig:mass_completeness} shows that our results are similar to the $95\%$ completeness limit obtained in \citetalias{Q1-SP031}. Figure~\ref{fig:mass_completeness} also shows the important improvement of about \mbox{$1 \,\rm dex$} in the $M_*$ completeness limit compared to results obtained for ELAIS-N1, which had the deepest optical/IR data of the three fields in \ac{lotss} Deep Fields DR1 \citep{Duncan2021}. Interestingly, Fig.~\ref{fig:mass_completeness} shows that \Euclid probes down to $M_*\approx7\times10^9 M_\odot$ at $z \approx 1.5$, which is about $1.5 \,\rm dex$ below the characteristic break in the stellar mass function \citep{McLeod2021TheSurveys}.

\section{\label{sc:RLAGN selection} Selecting radio AGN}
Having cross-matched \ac{lofar} sources with \Euclid in the \ac{edfn} and assessed contamination, we need to separate \ac{rlagn} from \ac{sfg}. Before proceeding, we clarify the terminology adopted throughout this paper. In the literature, radio-\ac{agn} are distinguished from \ac{sfg} using a variety of methods, including spectroscopy, extended radio emission, IR colours, and the detection of a radio excess \citep[e.g.,][]{Drake2024, Hardcastle2024}. In the following, we classify as radio-\ac{agn} sources showing radio emission significantly exceeding that expected from the host-galaxy \ac{sfr} \citep{Smith2021, Best2023, Shenoy2025}. We use the terms `radio excess', \ac{rlagn}, and radio-\ac{agn} interchangeably throughout. The underlying physical idea is that galaxy radio emission at frequencies \mbox{$\nu \lesssim 1 \,\rm{GHz}$} is dominated by non-thermal synchrotron radiation produced by highly energetic particles \citep{Condon1992RadioGalaxies}. These particles are accelerated either in supernova remnants of massive stars \citep{Blumenthal1970BremsstrahlungGases, KennicuttJr.1998TheGalaxies} or by \ac{agn} jets and outflows.

At 144 MHz, the emission from \ac{sfg} is therefore dominated by synchrotron radiation associated with star formation, leading to a tight correlation between \ac{sfr} and $L_{\rm 144}$ \citep{Wang2019ATracer}. Several studies of this correlation have found very similar parameters for a power-law relation \citep[e.g.,][]{Gurkan2018, Smith2021, Best2023}. There has been some discussion about whether to include a stellar-mass dependence \citep{Gurkan2018, Smith2021, Delvecchio2021The4}. This can be explained by the deeper gravitational potential of more massive galaxies allows them to retain their cosmic ray population for longer timescales, thus resulting in more radio emission for a given \ac{sfr}. Indeed, previous studies have shown that higher-mass galaxies exhibit higher $L_{144}$ at fixed \ac{sfr}, and that including $M_*$ reduces the scatter in the empirical relation (e.g., \citealt{Read2018TheLOFAR/H-ATLAS, Smith2021, McCheyne2022}). 

\cite{Shenoy2025} explored the dependence of the \mbox{$L_{144}$--\ac{sfr}} relation on several physical parameters, and found that $M_*$ cannot be neglected. Hence, we adopt their relation, i.e.
\begin{equation} \label{eq:SFR-L relation}
    L_{144}=L_c\,\left(\frac{\mathrm{SFR}}{M_\odot \, \rm{yr}^{-1}}\right)^\beta\,\left(\frac{M_*}{10^{10}M_\odot}\right)^\epsilon \,\rm W \, Hz^{-1}\; ,
\end{equation}

\noindent where $L_c$ is the normalisation and $\beta$ and $\epsilon$ are power law indices for the \ac{sfr} and $M_*$ dependence, respectively. We adopt the best-fit parameters found by \cite{Shenoy2025} for their photometric data, which are \mbox{$\log_{10}(L_c/\rm{W}\,\rm{Hz}^{-1}) = 21.848 \pm 0.003$}, $\beta = 1.107 \pm 0.008$, and $\epsilon=0.244\pm0.007$. 

To identify radio-excess sources, we must account for the intrinsic scatter that this relation exhibits \citep{Smith2021, Das2024}. Therefore, we require galaxies to lie at least $0.8 \,\rm dex$ (about $3\,\sigma$) above the mass-dependent $L_{144}$--$\rm{SFR}$ relation, consistent with previous studies \citep{Cochrane2023, Shenoy2025}. Because the \ac{sfr} estimates are subject to considerable uncertainty, we adopt a probabilistic approach. For each source, we draw $5\,000$ samples of $S_{144}$, \ac{sfr}, $M_*$, and $z$ from Gaussian distributions centred on their median values. The standard deviation of each distribution is set to the mean of the upper and lower $1 \,\sigma$ uncertainties. These uncertainties are taken either directly from the catalogue or are calculated using the differences between the 84th and 50th percentiles and between the 50th and 16th percentiles. For the sources with a spectroscopic redshift, we adopt a small uncertainty in redshift of \mbox{$\sigma_z = 0.001\,(1+z_{\rm spec})$}. We classify a source as having a radio excess if it exceeds the threshold value in at least $90\%$ of the realisations. This results in $2\,748$ out of $10\,794$ sources being classified as radio-excess.

\begin{figure}[htbp!]
    \centering
    \includegraphics[width=0.49\textwidth]{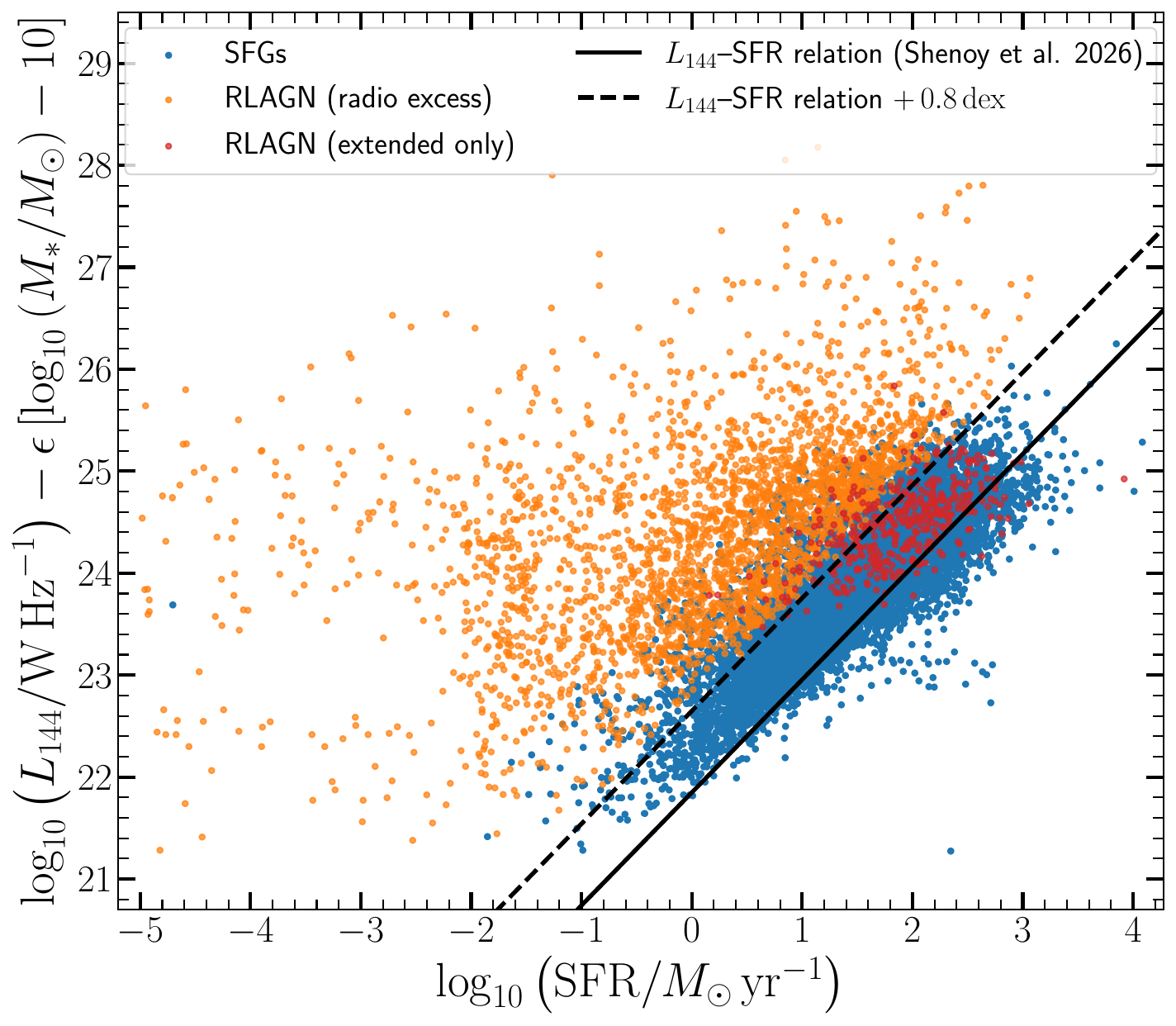}
    \caption{Radio luminosity $L_{\rm 144}$ as a function of \ac{sfr} and $M_*$ for all cross-matched sources: \ac{rlagn} with a radio excess (orange), extended \ac{rlagn} with no radio-excess (red), \ac{sfg} (blue). The solid black line shows the mass-dependent $L_{144}$--$\mathrm{SFR}$ relation from Eq.~\eqref{eq:SFR-L relation}, and the dashed black line shows the radio-excess threshold, using the median $M_*$ of our matched sample.
    }
    \label{fig:L144-SFR}
\end{figure}

In addition to the classification based on a radio excess, we also include all extended radio sources with a major axis size larger than $75 \,\rm kpc$. This value was chosen because an angular diameter distance of $75 \,\rm kpc$ spans at least $\ang{;;8.9}$ on the sky, which is resolvable with the \ang{;;6} resolution. Such extended radio emission is highly likely not powered by star formation, but by synchrotron emission from \ac{agn} jets. The full sample contains $710$ extended radio sources, of which $300$ are already included among the radio excess sources. This means that the full classification yields $3\,158$ \ac{rlagn} out of $10\,794$ radio sources, of which $1\,177$ objects have physical parameters from \citetalias{Q1-Laloux} and $1\,981$ from \citetalias{Q1-SP031}. In Fig.~\ref{fig:L144-SFR} we show \mbox{$\log_{10}(L_{\rm 144}/\mathrm{W}\,\mathrm{Hz}^{-1}) - \epsilon\left[\log_{10}(M_*/M_\odot) - 10\right]$} as a function of the \ac{sfr} for all matched radio sources in the \ac{edfn}, as well as the \mbox{$L_{144}$--$\rm{SFR}$} relation. The radio excess criterion of the \mbox{$L_{144}$--$\rm{SFR}$} relation plus \mbox{$0.8 \,\rm dex$} and the radio sources with an excess are also shown in Fig.~\ref{fig:L144-SFR}.

As discussed in Sect.~\ref{sc:Intro}, we define as \acp{herg} (\acp{lerg}) those \ac{rlagn} with evidence (absence of) of a radiative-mode \ac{agn}. Generally, a small subset of \ac{lofar} sources can show signs of a radiative-mode \ac{agn} (e.g., \citealt{Duncan2021, Kondapally2022}). As discussed in Sect.~\ref{sc:Intro}, spectroscopic data would allow us to more robustly distinguish between these two classes \citep{Arnaudova2025}. Although surveys such as DESI provide spectroscopic observations, they limit the coverage and analysis of radio sources in \ac{edfn} due to optical-based target selection and incompleteness in the optical selection function, especially beyond $z\sim0.6$. More complete spectroscopic observations of radio-selected sources are therefore needed to reliably distinguish between \acp{herg} and \acp{lerg}.

Instead, we make use of the \ac{agn} fraction $f_\mathrm{AGN}$ obtained from the \ac{sed} fitting with \ac{cigale} (\citetalias{Q1-Laloux}) to identify radiative-mode \ac{agn}. This \ac{agn} fraction is defined as the ratio of the \ac{agn} integrated luminosity to the total luminosity within the wavelength range of $5$--$20 \,\micron$ \citep{Thorne2022DeepFunction}. We adopt the method discussed in \citetalias{Q1-Laloux}, who found that the \ac{agn} fraction itself shows large uncertainties and instead defined a parameter \mbox{$f_\mathrm{AGN,\, low}\equiv \mathrm{max}(f_\mathrm{AGN}-\delta f_\mathrm{AGN},0)$} where \mbox{$\delta f_\mathrm{AGN}$} is the \mbox{$1\,\sigma$} uncertainty on $f_\mathrm{AGN}$. By comparing with a spectroscopic sample in the \ac{edfn}, they found that requiring \mbox{$f_\mathrm{AGN,\, low} > 0.2$} to select \ac{agn} leads to a relatively small contamination of about $9\%$\footnote{As a comparison, requiring \mbox{$f_\mathrm{AGN,\, low} > 0.0$} or \mbox{$f_\mathrm{AGN,\, low} > 0.075$} to select \ac{agn} leads to a contamination of $25\%$ and $15\%$, respectively.} based on the fraction of non-\ac{agn} sources being misclassified as \ac{agn} (\citealt{Q1-SP027}; \citetalias{Q1-Laloux}).

\begin{figure}[htbp!]
    \centering
    \includegraphics[width=0.49\textwidth]{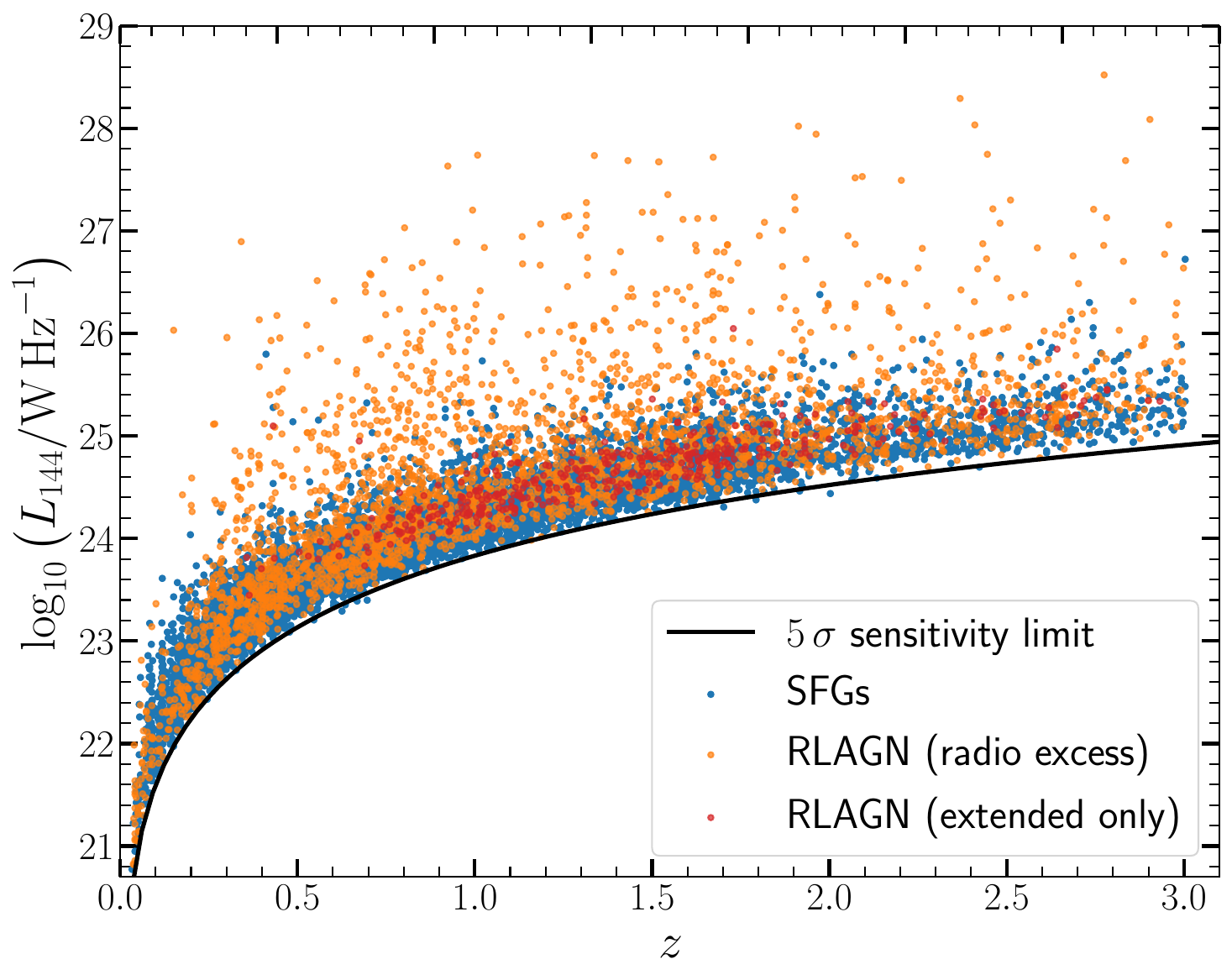}
    \caption{Radio luminosity $L_{\rm 144}$ as a function of redshift $z$. The colour scheme is the same as in Fig.~\ref{fig:L144-SFR}. The black line shows the sensitivity limit corresponding to an \ac{rms} flux of \mbox{$\sigma=32 \,\si{\uJy}\,\mathrm{beam}^{-1}$.}}
    \label{fig:L144-z}
\end{figure}

Using $f_\mathrm{AGN}$ obtained through \ac{sed} fitting with \ac{cigale}, we classified \ac{rlagn} as \acp{herg} for sources with $f_\mathrm{AGN,\, low}$ larger than $0.2$ and as \acp{lerg} otherwise. This selection resulted in a sample containing $353$ \acp{herg} and $2\,805$ \acp{lerg}. The resulting number of sources for every selection step discussed above is shown in Table~\ref{tab:source selections}. To illustrate the cross-matched catalogue, we show $L_{\rm 144}$ as a function of $z$ in Fig.~\ref{fig:L144-z}. We also show the sensitivity limit, assuming a spectral index $\alpha=-0.7$ and using a central point source $5\,\sigma$ detection limit of $32 \, \si{\uJy}\,\mathrm{beam}^{-1}$.

\begin{table}[!ht]
    \centering
    \caption{Summary of the number of sources in the \ac{lofar}-\Euclid cross-matched sample of the \ac{edfn} for different selection steps.}
    \begin{tabular}{cc}
        \hline\hline \noalign{\vskip 2pt} Selection step & Number of sources \\ \hline\noalign{\vskip 2pt}
        All radio sources in EDF-N & $23\,309$ \\
        Sources outside star mask & $19\,550$ \\
        Cross-matched sources & $14\,978$\\
        With physical parameters & $13\,477$\\
        With reliable physical parameters & $10\,794$\\
        \ac{rlagn} & $3\,158$ \\
        \acp{lerg} & $2\,805$ \\
        \acp{herg} & $353$ \\ \hline
    \end{tabular}
    \label{tab:source selections}
\end{table}

\section{\label{sc:RLAGN fractions} Incidence of LERGs}
With the cross-matched sample, the \ac{rlagn} selection, the $M_*$ completeness, and the cross-matching contamination all in place, we can now calculate the true \ac{rlagn} incidence as a function of $M_*$ and $z$ in the \ac{edfn}. Since the number of \acp{herg} is too limited for a detailed statistical analysis (see Table~\ref{tab:source selections}), we only calculate the \ac{lerg} incidence $f_{\rm LERG}$ in the following sections. To calculate $f_{\rm LERG}$, we follow the method of \cite{Janssen2012} to account for the flux-limited nature of the radio and optical samples. Because raw source counts are biased towards intrinsically luminous or lower-redshift objects, we weight each source by $V_{\mathrm{max}}$. This is the maximum co-moving volume over which a source could be detected given its luminosity and the survey sensitivity limits, following the $V_{\mathrm{max}}$ formalism \citep{Schmidt1968}. This method takes into account the redshift limits from both the radio and optical selection criteria, as well as the limits of the redshift range being analysed.

The maximum co-moving volume in which source $i$ would still be detected given the $5\,\sigma$ detection limit based on the local radio \ac{rms} level is denoted by $V_{\mathrm{max,}\, i}$. For the general galaxy sample within the \ac{lofar} footprint, we determine the maximum co-moving volume of source $j$ by using the $3\,\sigma$ detection limit for the VIS band. We calculate the volume-corrected $f_{\rm LERG}$ as a function of host properties $M_*$ and $L_{\rm 144}$ in different $z$ bins, using
\begin{equation}
    f_{\mathrm{LERG}}^{L,\, M}
    \;=\;
    \Bigg(\sum_{i \,\in\, R_{L,\, M}} \frac{1}{C_{\rm ID} V_{\max,\,i}}\Bigg)
    \Bigg(\sum_{j \,\in\, G_{M}} \frac{1}{V_{\max,\,j}}\Bigg)^{-1}\; ,
    \label{eq:frac}
\end{equation}
where $R_{L,\, M}$ is the number of \acp{lerg} in a given $M_*$ and $L_{\rm 144}$ bin and $G_{M}$ are all galaxies in a given $M_*$ bin. We define $C_\mathrm{ID}$ as a correction factor to account for the incompleteness of optical identifications for the radio sources. We determine this factor by evaluating the ratio of the number of radio sources with an optical counterpart to the total number of radio sources outside the star mask within a given radio flux bin. We only consider $M_*$ bins that are above the $M_*$ completeness limit discussed in Sect.~\ref{sub-sc: Mstar completeness}. Uncertainties on $f_{\rm LERG}$ are estimated assuming Poisson statistics \citep{Wall2003PracticalAstronomers}, and $f_{\rm LERG}$ is only calculated when there are more than five sources in a bin to avoid small-number fluctuations. As discussed in Sect.~\ref{sub-sc: matching contamination}, we need to account for the increase in cross-matching contamination as $M_*$ decreases. Therefore, we interpolated $f_{\rm true}(M_*)$ shown in Fig.~\ref{fig:contamination fraction} and used this to correct $f_{\rm LERG}$, i.e., \mbox{$f_\mathrm{LERG}\left(M_*\right)=f_\mathrm{LERG,\, meas.}\left(M_*\right) \times f_\mathrm{true}\left(M_*\right)$}. Similarly, the total quadratic uncertainty on $f_\mathrm{LERG}\left(M_*\right)$ is defined as the quadratic sum of the Poisson uncertainty on $f_\mathrm{LERG,\, meas.}\left(M_*\right)$ and the uncertainty on $f_\mathrm{true}\left(M_*\right)$.

\begin{figure*}
    \centering
    \includegraphics[width=\textwidth]{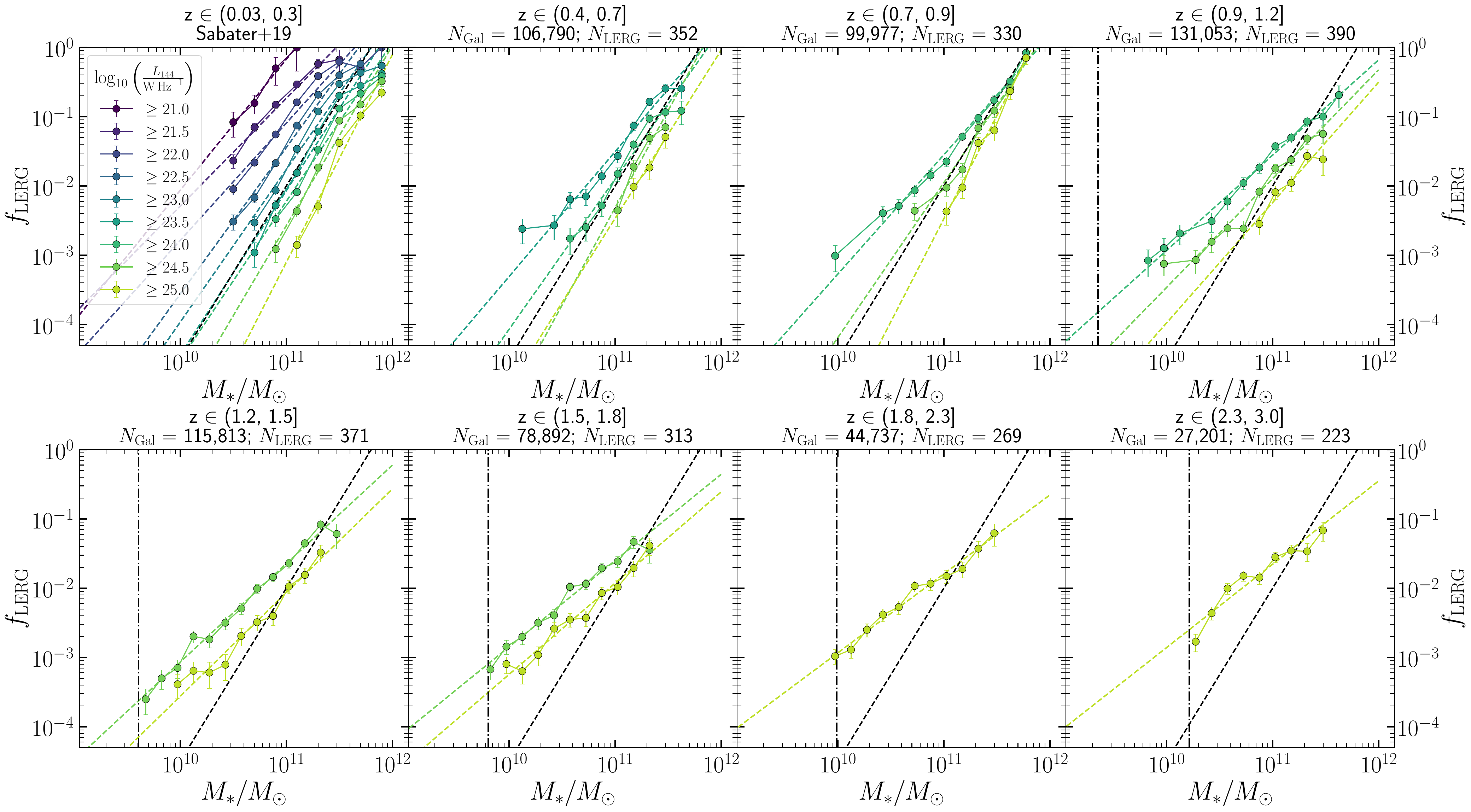}
    \caption{Fraction $f_\mathrm{LERG}$ as a function of $M_*$ for different $L_{\rm 144}$ limits and its redshift evolution in the range \mbox{$0.03\leq z \leq 3.0$}. The dot-dashed vertical lines indicate the $M_*$ completeness limit for each redshift bin. The black dashed line in each panel shows the \mbox{$f_\mathrm{RLAGN}= 0.01\,(M_*/10^{11}M_\odot)^{2.5}$} relation found in the local Universe \citep{Best2005a}. The coloured dashed lines are the power-law fits of $f_\mathrm{LERG}$ for every radio luminosity limit that is detectable in each panel. \emph{Top left}: Fraction $f_\mathrm{RLAGN}$ for the local Universe using \ac{lotss} DR1 data \citep{Sabater2019}.}
    \label{fig:AGN fractions}
\end{figure*}

As discussed above, we first binned all \acp{lerg} by redshift to apply a volume correction. For this, we adopted the redshift values \mbox{$z\in\curly{0.4, 0.7, 0.9, 1.2, 1.5, 1.8, 2.3, 3.0}$}. The lower limit \mbox{$z=0.4$} was chosen to ensure a large enough cosmological volume is observed with the $8.73\,\mathrm{deg}^2$ sky area. For the local Universe ($z<0.3$), the sky area is too small to detect a significant number of sources for an analysis of the \ac{lerg} incidence with both $L_{144}$ and $M_*$. Instead, we will use $f_\mathrm{RLAGN}$ found by \cite{Sabater2019}. This work used the \ac{lotss} DR1 data \citep{Shimwell2019}, which covers $424 \,\mathrm{deg}^2$ in the \ac{hetdex} field, to calculate the \ac{rlagn} incidence to $z=0.3$. The different radio-\ac{agn} selection and higher radio luminosities probed by \cite{Sabater2019} do mean that their sample contained a non-negligible sample of \acp{herg}. Nevertheless, the large sky area serves as a good benchmark for the local Universe.

We binned $M_*$ in the range \mbox{$9.15 \leq \log_{10}(M_*/M_\odot) \leq 11.85$} in steps of \mbox{$0.15 \,\rm dex$}. This step size was chosen because the $M_*$ values in our sample have an average uncertainty of \mbox{$0.1 \,\rm dex$}. The $L_{\rm 144}$ limits were chosen to be \mbox{$\log_{10}(L_{144}/\,\rm{W}\,\rm{Hz}^{-1}) \in \curly{23.5,24.0,24.5,25.0}$}. The lower $L_{\rm 144}$ limit matches the $\geq5\,\sigma$ detection limit in the deepest regions of the radio data in the \ac{edfn} at $z=0.4$ \citep{Bondi2024}. The highest $L_{\rm 144}$ limit ensures at least one $L_{144}$ bin in every redshift bin, as shown in Fig.~\ref{fig:L144-z}.

Using this binning scheme and the method described above, the resulting $f_\mathrm{LERG} (M_*,\, z)$ are shown in Fig.~\ref{fig:AGN fractions}. Each panel represents a $z$ bin, with fractions plotted at the centres of the $M_*$ bins. As a reference, we also show $f_\mathrm{RLAGN}$ values from \cite{Sabater2019} in the top left panel, as well as the relation \mbox{$f_\mathrm{RLAGN}= 0.01\,(M_*/10^{11}M_\odot)^{2.5}$} for the local Universe \citep{Best2005a}. This relation was derived for radio-\ac{agn} with radio luminosities at \mbox{$1.4 \,\rm GHz$} $L_{1.4} > 10^{23}\,\rm{W}\,\rm{Hz}^{-1}$ and redshifts $0.03 \leq z \leq 0.3$. Assuming a spectral index of $\alpha=-0.7$, this corresponds to \mbox{$\log_{10}(L_{144}/\rm{W}\,\rm{Hz}^{-1}) \geq 23.7$}. The results reveal a strong correlation between $f_\mathrm{LERG}$ and $M_*$ over a wide redshift range extending beyond the epoch of cosmic noon. Previous studies have found this dependence regardless of the \ac{rlagn} selection method (e.g., \citealt{Williams2015, Sabater2019, Kondapally2022, Kondapally2025}).

\subsection{\label{sub-sc:RLAGN fractions evolution} Redshift evolution of the LERG incidence}
Qualitatively, Fig.~\ref{fig:AGN fractions} shows that the \ac{lerg} incidence resembles the relation observed in the local Universe to $z = 0.7$, before flattening towards higher $z$. The fraction $f_\mathrm{LERG}$ at fixed $M_*$ increases with $z$, implying that \acp{lerg} are dominated by galaxies spanning a broader $M_*$ range at earlier epochs. For the highest stellar masses, $M_* \geq 10^{11.25}\, M_\odot$, $f_\mathrm{LERG}$ approaches $10\%$ and exhibits only a modest increase with $z$, implying that the flattening of the power-law relation is driven mainly by the stronger evolution of lower-mass galaxies. Both the increase in normalisation and the flattening are consistent with previous studies \citep{Williams2015, Wang2024Cosmic4, Kondapally2025}. Interestingly, Fig.~\ref{fig:AGN fractions} also suggests a slight deviation from the approximately linear relation at lower $M_*$ in several $z$ bins. Although the current results do not allow us to quantify this deviation robustly, it may indicate a change in the dominant \ac{smbh} triggering mechanisms in lower-mass galaxies. 

In the redshift range $z = 0.7$--$0.9$, there is an indication that almost all the most massive galaxies \mbox{($M_* > 10^{11.7}\, M_\odot$)} with \mbox{$\log_{10}(L_{144}/\rm{W}\,\rm{Hz}^{-1}) \geq 24.0$} host an \ac{lerg}, consistent with \cite{Sabater2019}. However, the number of galaxies with \mbox{$M_* > 10^{11.7}\, M_\odot$} is small, with about $10$ objects per redshift bin. Our requirement of at least five radio sources per bin means most of the $f_\mathrm{LERG}$ values at those masses are not shown. These small-number statistics are more prone to statistical fluctuations, meaning that we would need larger samples to study whether $f_\mathrm{LERG}$ reaches $100\%$ in more $z$ bins.

\begin{figure}
    \centering
    \includegraphics[width=0.49\textwidth]{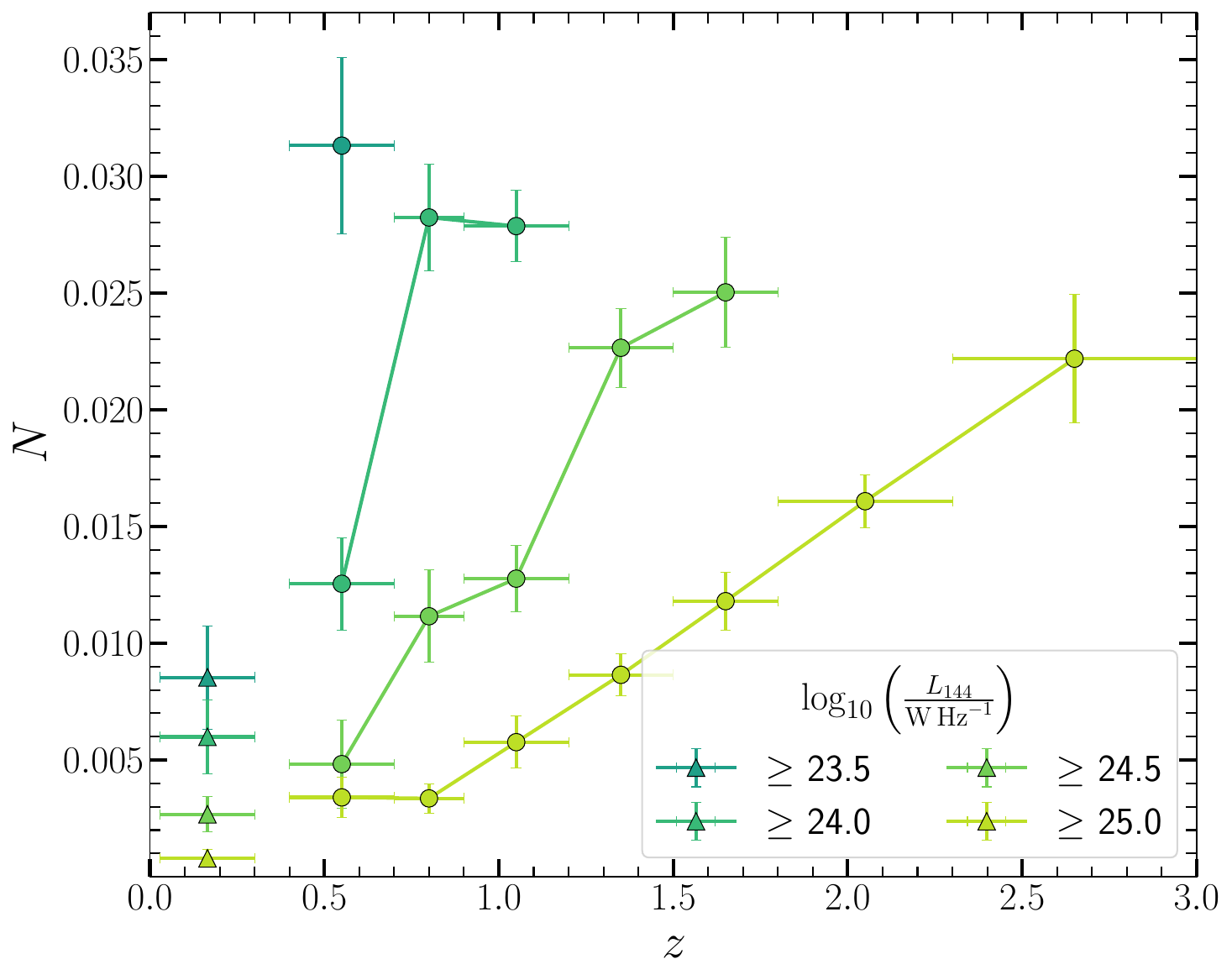}
    \caption{Redshift evolution of normalisation $N$ for the powerlaw fits of $f_\mathrm{LERG}$. Horizontal uncertainties indicate the width of the redshift bins, and the different colours represent different $L_{\rm 144}$ limits. Triangles represent $N$ obtained from fitting $f_\mathrm{RLAGN}$ in the local Universe \citep{Sabater2019}, while the circles show $N$ for $f_\mathrm{LERG}$ in the \ac{edfn}.}
    \label{fig:power law norm evolution}
\end{figure}

The strong dependence of $f_\mathrm{LERG}$ on $M_*$ may reflect the fuelling mechanisms operating in \acp{lerg} \citep[e.g.,][]{Allen2006TheGalaxies, Hardcastle2007HotGalaxies}. \acp{lerg} are generally associated with radiatively inefficient accretion of hot gas, which is expected to be most effective in massive galaxies with substantial hot gaseous haloes. This gas reservoir may explain the higher incidence of \acp{lerg} in more massive galaxies. This is consistent with the balance between radio-source heating and radiative cooling losses proposed by \cite{Best2006AGN-controlledGalaxies}, in which the hot halo persistently fuels the central \ac{smbh}, while \ac{agn} activity offsets cooling. At lower stellar masses ($M_* \lesssim 10^{11}\, M_\odot$), galaxies are more frequently disc-dominated and host lower-mass \acp{smbh}, potentially hindering sustained hot-mode accretion. Together, these effects may explain the observed increase of $f_\mathrm{LERG}$ with $M_*$, although our observations alone cannot establish the origin of this steep dependence.

\begin{figure}
    \centering
    \includegraphics[width=0.49\textwidth]{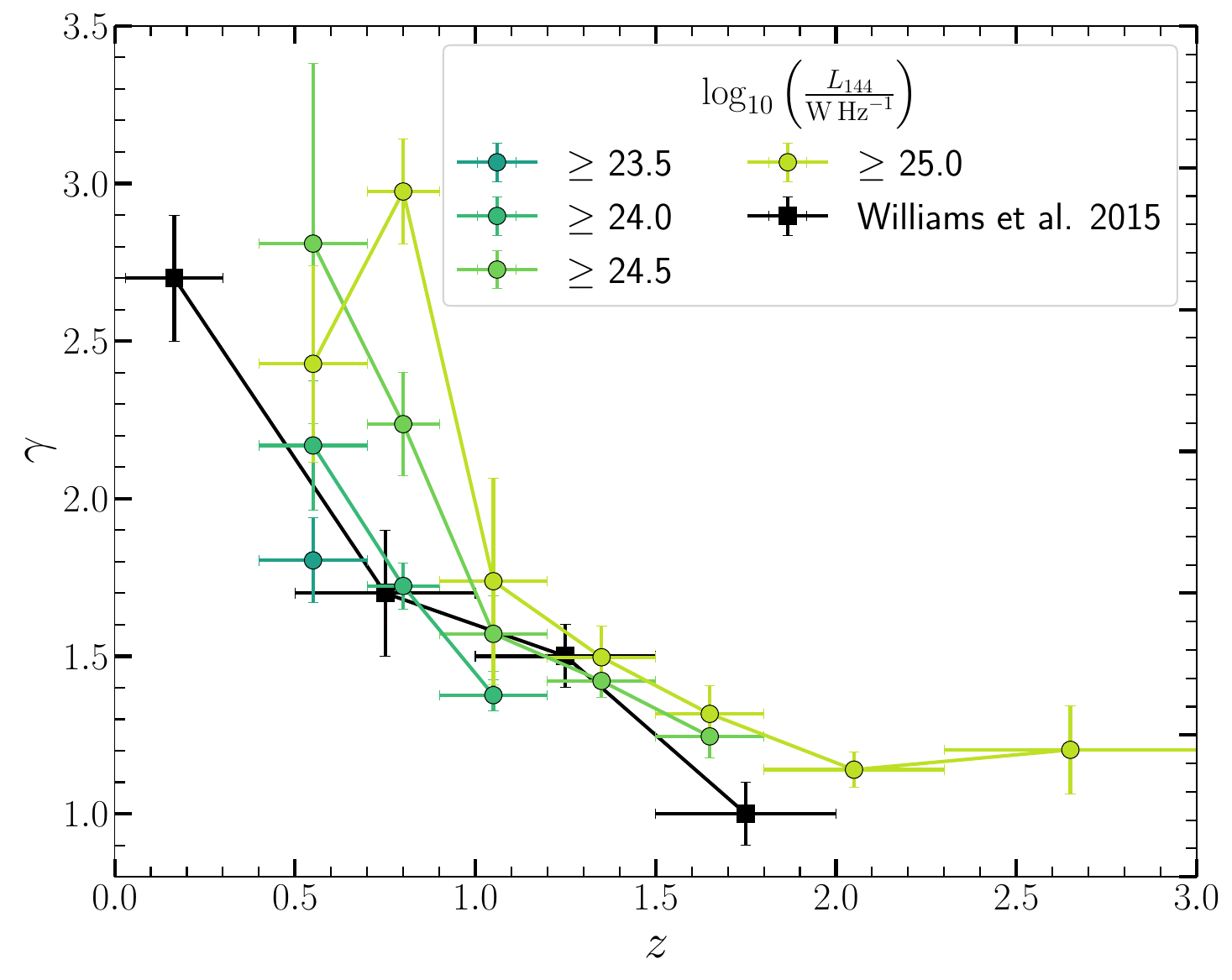}
    \caption{Same as in Fig.~\ref{fig:power law norm evolution}, but for $\gamma$. The black squares show $\gamma$ found by \cite{Williams2015} for a combined sample of \acs{sdss}-\acs{nvss} in the local Universe and \ac{vla}-COSMOS for higher redshifts.}
    \label{fig:power law index evolution}
\end{figure}

To quantify the redshift evolution of the $f_\mathrm{LERG}$--$M_*$ relation, we fit a power law in each $z$ bin. For a normalisation $N$ at \mbox{$M_*=10^{11}M_\odot$} and a power-law index $\gamma$, the relation takes the form
\begin{equation}\label{eq:powerlaw}
    f_\mathrm{LERG} = N \,\left(\frac{M_*}{10^{11}M_\odot}\right)^\gamma\; .
\end{equation}
We fit using \acl{odr}\footnote{\url{https://docs.scipy.org/doc/scipy/reference/odr.html}} (\acs{odr}\acused{odr}, \citealt{Boggs1990OrthogonalRegression}), requiring measurements across at least three $M_*$ bins for a given $L_{\rm 144}$ limit. We adopt uncertainties of $0.15\,\rm dex$ in \mbox{$\log_{10}(M_*/M_\odot)$}, corresponding to the bin width, and Poissonian uncertainties in $f_{\rm LERG}$. We also fit $f_\mathrm{RLAGN}$ from \cite{Sabater2019} for comparison with the local Universe. The fits are also shown in Fig.~\ref{fig:AGN fractions}, and the evolution of the fitted power-law parameters with redshift is shown in Figs.~\ref{fig:power law norm evolution} and~\ref{fig:power law index evolution}.

Focusing on the results for $N$ in Fig.~\ref{fig:power law norm evolution}, we see that the normalisation within a given $z$ bin decreases as we set higher $L_{\rm 144}$ limits. This decrease is to be expected because more luminous radio sources are intrinsically rarer. Interestingly, there is a very strong increase in the overall \ac{lerg} incidence with $z$ compared to the local Universe. The overall incidence of \acp{lerg} increases for all $L_{\rm 144}$ limits, and especially the most luminous objects show an increase by a factor of up to $20$ from the first to the last $z$ bin. The incidence of less luminous sources also increases substantially, but the current depth of the radio data cannot definitively determine whether they will show similar growth to $z=3.0$. The increasing normalisation for the most luminous sources has been found before (e.g., \citealt{Williams2015}), and supports the idea that radio-\ac{agn} activity was more prevalent at higher redshifts (e.g., \citealt{Heckman2014, Alexander2025WhatProgress}).

\begin{figure*}
    \centering
    \includegraphics[width=0.83\textwidth]{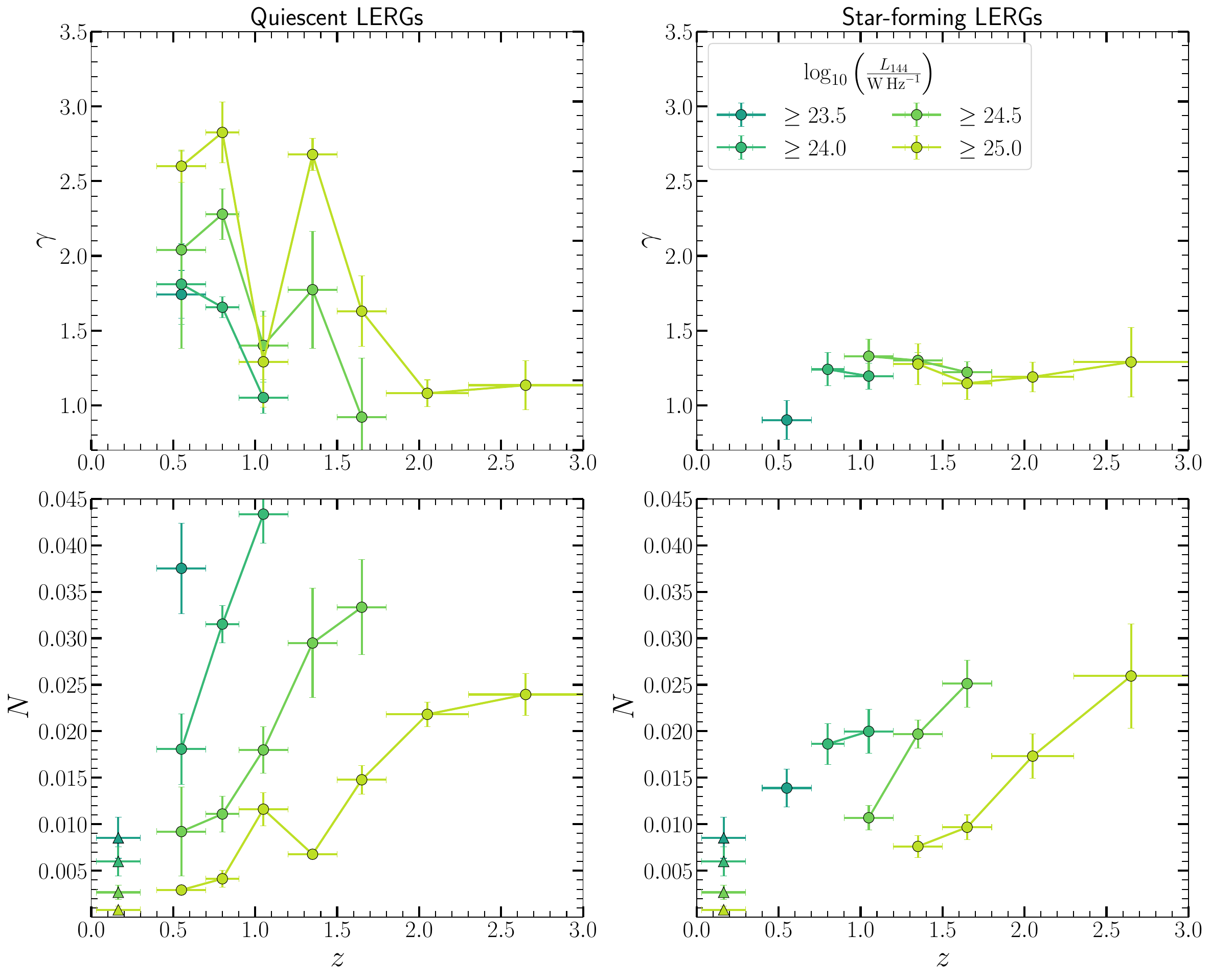}
    \caption{Same as in Figs.~\ref{fig:power law norm evolution} and \ref{fig:power law index evolution}, but for quiescent and star-forming \acp{lerg}. Quiescent galaxies were classified using $\mathrm{sSFR} \leq f_\mathrm{sSFR}/t_\mathrm{H}(z)$ with $f_\mathrm{sSFR}=1/3$. \emph{Left}: Results for quiescent \acp{lerg} where the upper panel shows the evolution of $\gamma$ and the lower panel shows the evolution of $N$. \emph{Right}: Results of power-law fits for star-forming \acp{lerg}.}
    \label{fig:evolution star-forming vs quiescent}
\end{figure*}

In Fig.~\ref{fig:power law index evolution}, we show the power-law index $\gamma$ for each $L_{\rm 144}$ limit and its evolution with $z$. This shows that $\gamma$ decreases with $z$ for all $L_{\rm 144}$ limits. For the highest radio luminosities, the index is $\gamma\approx2.5$ up to $z=1.0$ and decreases to $\gamma \leq 1.5$ towards the highest redshifts. This behaviour is consistent with results obtained for the COSMOS field \citep{Williams2015, Wang2024Cosmic4} and the \ac{lofar} Deep Fields \citep{Kondapally2022}, where slopes of \mbox{$\gamma \simeq 1.0$--$1.5$} are found at \mbox{$z \gtrsim 1.5$}. For example, the evolution reported by \cite{Williams2015} for sources with \mbox{$L_{1.4} > 10^{24}\,\rm{W}\,\rm{Hz}^{-1}$} is remarkably similar to our results for \mbox{$\log_{10}(L_{144}/\rm{W}\,\rm{Hz}^{-1}) \geq 24.5$}. They interpreted this flattening as evidence for an increasing contribution from HERG activity or, more generally, cold-mode accretion at higher $z$. While their sample could not be separated into \acp{herg} and \acp{lerg}, we excluded \acp{herg} using the \ac{cigale}-derived $f_\mathrm{AGN}$ parameter (see Sect.~\ref{sc:RLAGN selection}). The persistence of the same trend in our sample therefore implies that this evolution also applies to the \ac{lerg} population and cannot be attributed solely to an increase in \ac{herg} activity. The flattening of the $f_\mathrm{LERG}$--$M_*$ relation, in particular at lower $L_{144}$ and $M_*$ where \ac{sfg} are expected to dominate the galaxy stellar mass function, suggests the presence of a star-forming \ac{lerg} population with a potentially different fuelling mechanism \citep{Kondapally2022, Kondapally2025}. We further investigate this in Sect.~\ref{sub-sc:SF vs quiescent} by studying the redshift evolution of $f_\mathrm{LERG}$ as a function of both $M_*$ and \ac{sfr}.

As discussed in Sect.~\ref{sub-sc:Final sample}, our \ac{edfn} cross-matched sources have physical parameters from \citetalias{Q1-SP031} or \citetalias{Q1-Laloux}. To test for potential biases from combining the catalogues, we performed two tests. First, for the $947$ \ac{rlagn} in both catalogues, we compared the \ac{pdf} of the photometric $z$, SFR, and $M_*$ to identify any discrepancies. Second, we calculated $f_\mathrm{LERG}$ as a function of $M_*$ and $L_{\rm 144}$ in the range $0.7\leq z \leq0.9$ separately, using $M_*$ from both catalogues to check for any systematic offset. Both tests showed that the physical parameters from both catalogues yield comparable results. More details about these tests can be found in Appendix~\ref{app:compare sed fitting}. 

We also repeated the analyses shown in Figs.~\ref{fig:AGN fractions}, \ref{fig:power law norm evolution}, and \ref{fig:power law index evolution} excluding extended sources without a radio excess. The results were consistent within uncertainties and retained the decreasing trend of $\gamma$ with redshift. Finally, we applied a more conservative \ac{herg} selection by requiring $f_\mathrm{AGN,\,low}>0.0$, which yielded $662$ \acp{herg} and $2\,496$ \acp{lerg}. Given its relatively high contamination rate \citep{Q1-SP027}, this threshold provides a conservative test of our \ac{herg} selection. The results were again consistent within uncertainties, showing evolution from $\gamma\approx2.5$ in the local Universe to $\gamma\approx1.0$ at higher redshifts. Together, these tests demonstrate the robustness of our results with respect to our classifications.

\subsection{\label{sub-sc:SF vs quiescent} Redshift evolution of star-forming and quiescent LERGs}
As mentioned in Sect.~\ref{sub-sc:RLAGN fractions evolution}, the flattening of the relation between $f_\mathrm{LERG}$ and $M_*$ with $z$ for \acp{lerg} has been observed in other works as well. For example, \cite{Kondapally2022, Kondapally2025} argued that this can be explained by both an increasing number of active \ac{sfg} and an increasing $f_\mathrm{LERG}$ among the star-forming population at fixed $M_*$ towards higher $z$. In particular, \cite{Kondapally2022} showed that more quiescent \acp{lerg} have a steep power-law dependence with a slope \mbox{$\gamma\approx2.5$}, while star-forming \acp{lerg} show \mbox{$\gamma\approx1.5$} out to \mbox{$z=1.5$}. 

Figure~\ref{fig:power law index evolution} shows that the flattening of $f_\mathrm{LERG}$ continues towards higher redshift down to a power-law index of $\gamma\approx1$ for the highest radio luminosities. An increasing fraction of star-forming relative to quiescent \acp{lerg} would naturally result in a flattening of $f_\mathrm{LERG}$ with $M_*$. An explanation for this is that the higher gas fractions at higher redshifts \citep[e.g.,][]{Genzel2015COMBINEDMASS, Tacconi2018PHIBSS:Fractions} result in an increasing number of \ac{sfg}. Moreover, at higher redshifts, galaxies have had less time to undergo the morphological transformation associated with mergers, resulting in a higher prevalence of disc-dominated systems compared to the local Universe \citep{Conselice2014, Martin2018}.

To test whether there is a difference between star-forming and quiescent \acp{lerg}, we split our sample of \acp{lerg} into two groups. The sub-classification is based on the idea that while \ac{sfg} follow the `main-sequence' of star formation in the \ac{sfr}--$M_*$ plane \citep[e.g.,][]{Whitaker2012THE2.5, Speagle2014A06, Schreiber2015TheDay}, quiescent galaxies lie below this sequence and we can use the \mbox{\acl{ssfr} ($\mathrm{sSFR}\equiv \mathrm{SFR}/M_*$)\acused{ssfr}} to distinguish between star-forming and quiescent galaxies \citep{Fontana2009TheUniverse, Pacifici2016THEGALAXIES, Merlin2018ChasingGOODS-South}. We classify a source as quiescent when it satisfies \mbox{$\mathrm{sSFR} \leq f_\mathrm{sSFR}/t_\mathrm{H}(z)$}, where $t_\mathrm{H}(z)$ is the age of the Universe at a certain $z$ in Gyr and $f_\mathrm{sSFR}=1/3$ defines the threshold. This threshold, adopted from \cite{Kondapally2022}, roughly divides our \ac{lerg} sample in half and is consistent with traditional $UVJ$ rest-frame colour-selected samples \citep{Pacifici2016THEGALAXIES, Carnall2020Timing5}. The classification results in $1\,535$ star-forming ($54.7\%$) and $1\,270$ quiescent ($45.3\%$) \acp{lerg}. To check our selection method, we also used the \mbox{$\rm NUV$--$r^+$} and \mbox{$r^+$--$J$} colours, which were shown to be consistent, as discussed in Appendix~\ref{app: quiet/sf galaxies}.

Following the procedure adopted earlier for the full \ac{lerg} sample, we calculate $f_\mathrm{LERG}$ as a function of $M_*$ for different $L_{\rm 144}$ limits in each redshift bin and fit a power law. For quiescent \acp{lerg}, the fraction is calculated relative to the quiescent galaxy population, and analogously for star-forming \acp{lerg}. Of the $1\,945\,312$ \Euclid\ galaxies within the \ac{lofar} \ac{edfn} footprint, $173\,479$ are classified as quiescent and $1\,771\,842$ as star-forming based on their \ac{ssfr}. The power-law fits to the star-forming and quiescent \ac{lerg} fractions are presented in Appendix~\ref{app: quiescent and sf lerg fractions}, and the redshift evolution of the fitted normalisation and slope is shown in Fig.~\ref{fig:evolution star-forming vs quiescent}. The power-law normalisations derived from \cite{Sabater2019} are included for comparison only, since they did not distinguish between star-forming and quiescent galaxies.

Despite the larger uncertainties arising from the smaller sample, star-forming \acp{lerg} exhibit significantly flatter power-law relations than quiescent \acp{lerg}. However, we do not find that the power-law index for quiescent sources remains $\gamma\approx2.5$ for the entire redshift range. Out to $z=1.5$, the slopes remain consistent with the results in \cite{Kondapally2022} for the highest $L_{144}$, but the slope decreases and becomes more similar to the star-forming population with $\gamma\approx1.5$ at higher redshifts. One caveat is that with the current depth of the \ac{lofar} data in the \ac{edfn}, we cannot probe lower radio luminosities across the entire redshift range as was done in \cite{Kondapally2022}. Nevertheless, the similar slopes at higher redshifts suggest that quiescent and star-forming systems may be triggered similarly at those redshifts, at least for the highest radio luminosities.

An explanation for this could be that the large availability of cold gas at high $z$ enables very efficient fuelling of the \ac{smbh}. However, once the cold gas reservoirs start to deplete and star formation stops, the fuelling process becomes dependent on either gas-rich mergers, which ignite star formation and drive cold gas to the centre of the galaxy, leading to $f_\mathrm{LERG}\propto M_*^{1.5}$. Alternatively, the fuelling will depend on the ability of the galaxy to accrete hot gas from the halo, which follows $f_\mathrm{LERG}\propto M_*^{2.5}$. This suggests an evolutionary picture where the higher availability of cold gas and higher merger rates at earlier epochs influence the \ac{lerg} incidence (\citealt{Hopkins2006ASpheroids, Chiaberge2015RADIOMERGERS}). 

Nevertheless, our study is currently limited by the lack of spectroscopic data for the full \ac{rlagn} sample, which would provide accurate redshifts and reliable classifications of \acp{lerg} and \acp{herg}. Combining spectroscopy with deeper radio data across wider areas is needed to obtain enough statistics to test whether the $f_\mathrm{LERG}$--$M_*$ relation persists to higher redshifts for less luminous radio sources.

\section{\label{sc:Conclusions}  Conclusions}
In this work, the new Euclid Q1 data in the \ac{edfn} have been combined with \ac{lofar} radio observations at $144 \,\mathrm{MHz}$ in the same field. The \ac{lofar} sources were cross-matched with optical/\ac{nir} Q1 sources using the \ac{lr} method based only on coordinates to obtain a highly reliable sample. We avoid the use of colours to prevent biasing our matching with any priors on galaxy colour. Subsequently, physical parameters for the cross-matched sample were obtained from the catalogues presented in \citetalias{Q1-SP031} and \citetalias{Q1-Laloux} for galaxies and \ac{agn} candidates, respectively. After correcting for $M_*$ incompleteness and cross-matching contamination, sources were classified into \ac{sfg} and \ac{rlagn}. The latter was done by identifying extended radio emission or a radio excess based on the \ac{sfr} and $L_{\rm 144}$ using a probabilistic approach. Finally, we classified all radio-\ac{agn} as \acp{lerg} or \acp{herg} using the \ac{agn} fraction obtained through \ac{sed} fitting.

The resulting \ac{lerg} incidence shows a tight correlation over 2 orders of magnitude in $M_*$ across a wide range of redshifts. Our results confirm that \acp{lerg} are preferentially associated with the most massive galaxies, as found before by \cite{Sabater2019}. Fitting the \ac{lerg} fractions with a power law showed that the relations flatten with redshift from $\gamma\approx2.5$ in the local Universe to $\gamma\approx1.0$ at $z=3.0$. Moreover, the normalisation of $f_\mathrm{LERG}$ was found to increase with redshift for all probed radio luminosities, with the highest radio luminosities $L_{144} \geq 10^{25}\,\rm{W}\,\rm{Hz}^{-1}$ increasing the most from the local Universe to $z=3.0$. Overall, our results show that the \ac{lerg} population becomes more dominated by less massive sources and that the incidence of \acp{lerg} increases significantly with increasing redshift.

One possible reason for the increased incidence at higher redshift could be the greater availability of cold gas, which may enable efficient accretion and thus more activity of the \ac{smbh} \citep{Kondapally2022, Kondapally2025, Q1-SP065}. In addition, the increasing dominance of less massive sources may be related to higher merger rates with increasing redshift. Given that the overall number density of less massive galaxies is higher than for the most massive objects, the \ac{rlagn} incidence could consequently be boosted by higher merger rates. In our analysis, we cannot disentangle the effects of the evolving merger rate and the increasing availability of cold gas. The observed flattening of $f_\mathrm{LERG}$ may therefore reflect a combination of these effects, although their relative contributions remain uncertain.

Separating the \ac{lerg} population into quiescent and star-forming sources suggested an evolutionary picture with particularly distinct behaviour in each class. In agreement with previous work \citep{Kondapally2022, Kondapally2025}, we find that quiescent \acp{lerg} show a significantly steeper $M_*$ dependence with $\gamma\approx2.5$ than star-forming \acp{lerg}, which show $\gamma\approx1.5$ for $z<2.0$. However, at the highest redshifts ($z\geq2.0$), both populations show similar relations with $\gamma\approx1.0$. Our interpretation is that the large abundance of cold gas at higher redshifts enables efficient fuelling of the \acp{smbh}. However, as cold gas becomes more scarce at lower redshifts, the fuelling mechanism changes to the accretion of hot gas. This phenomenon is more dependent on the mass of the host galaxy and its halo. In this picture, galaxy mergers can resupply the availability of cold gas to trigger star formation and efficient fuelling of the \ac{smbh} for some galaxies. Other sources that do not undergo mergers become quiescent, and their host mass influences how efficiently they can accrete hot gas and thus whether an \ac{rlagn} can be triggered.

An interesting avenue to explore would be to check how many of our classified \acp{lerg} are part of a known galaxy group or cluster. The \ac{rlagn} incidence could be linked to the host galaxy environment, which is often seen as one of the main factors determining the \ac{herg}/\ac{lerg} appearance and \ac{rlagn} incidence in general \citep{Heckman2014, Koulouridis2024AGNsState, Hale2025TheFields}. Galaxy mergers are also thought to play an important role in triggering \ac{agn} and leading to radio activity \citep{Q1-SP065}. However, the number of \ac{rlagn} in the \ac{edfn} data is too low to estimate the \ac{lerg} incidence robustly while incorporating these environmental parameters. This would be something to explore further in the first large data release of \Euclid (DR1).

Future studies of \acp{rlagn} and their hosts will benefit from combining the wide \Euclid survey with \ac{lotss} DR3 \citep{Shimwell2026TheRelease} and other large radio surveys. Although \ac{lotss} DR3 is shallower than the \ac{edfn}, and thus less suited to studies extending to $z=3.0$, its larger sample will enable more robust statistical analyses at lower redshifts. This will allow studies to consider not only redshift and $M_*$, but also other properties like the \ac{herg}/\ac{lerg} classification, host-galaxy morphology and environment. Moreover, the upcoming spectroscopic WEAVE-\ac{lofar} survey \citep{Smith2016} will further improve redshift quality and source classification of \ac{lofar} radio sources. This survey is planned to obtain a million spectra for sources detected across \ac{lotss} using the \acl{wht} Enhanced Area Velocity Explorer (WEAVE)\acused{weave} multi-object spectrograph \citep{Jin2023}. In addition, the final \ac{lotss} Deep Fields (e.g., \citealt{deJong2024, Shimwell2025TheField}) will provide detailed sub-arcsecond resolution imaging over an estimated $60 \,\mathrm{deg}^2$.\footnote{The final area covered by each \ac{lofar} Deep Field will depend on the exact layout of the central and neighbouring pointings, as well as the primary-beam cuts in the processing.} Comparable angular resolutions will improve cross-matching reliability, enabling direct comparisons with optical surveys such as \Euclid and providing a powerful resource for studying the evolution of radio-\ac{agn} and \ac{agn} feedback.

\begin{acknowledgements}
SV thanks C. Saulder for compiling the external spectroscopic redshift catalogues for the Q1 data and making this available to the Euclid Consortium. SV acknowledges support from NWO through The Dark Universe Science Collaboration (DUSC): Cosmology with the \Euclid mission (OCENW.XL21.XL21.025; Grant ID: \url{https://doi.org/10.61686/NHOQI85483}). 
RK acknowledges support from the Leverhulme Trust through the Leverhulme Early Career Fellowship.
LKM is grateful for support from a UKRI FLF [MR/Y020405/1] and LOFAR-UK via STFC [ST/V002406/1]. 
ALM acknowledges support through the European Space Agency (ESA) Research Fellowship Programme.
GJW expresses grateful thanks for the award of an Emeritus Fellowship from The Leverhulme Trust.
LOFAR data products were provided by the LOFAR Surveys Key Science project (LSKSP; \url{https://lofar-surveys.org/}) and were derived from observations with the International LOFAR Telescope (ILT). LOFAR \citep{VanHaarlem2013} is the Low Frequency Array designed and constructed by ASTRON. It has observing, data processing, and data storage facilities in several countries, which are owned by various parties (each with their own funding sources), and which are collectively operated by the ILT foundation under a joint scientific policy. The efforts of the LSKSP have benefited from funding from the European Research Council, NOVA, NWO, CNRS-INSU, the SURF Co-operative, the UK Science and Technology Funding Council, and the Jülich Supercomputing Centre.
\AckEC
\AckQone
\AckDatalabs
\end{acknowledgements}

\bibliography{Euclid, RLAGN_paper_specific}

\clearpage
\begin{appendix}
\section{Comparison of LERG fractions for physical parameters obtained with diﬀerent methods}\label{app:compare sed fitting}
As discussed in Sect.~\ref{sub-sc:RLAGN fractions evolution}, we make use of two catalogues to provide physical properties and photometric redshifts for the \Euclid sources. For the analysis in this paper, it is important to verify that this will not introduce bias into the calculated \ac{lerg} incidence. Therefore, we look at all objects for which we have parameter estimates from both \citetalias{Q1-SP031} and \citetalias{Q1-Laloux}. Of the $10\,794$ radio sources with reliable physical parameters in the final catalogue discussed in Sect.~\ref{sub-sc:Final sample}, $3\,653$ are present in both catalogues. The first test is to compare the \acp{pdf} of the photometric redshift, \ac{sfr}, and $M_*$ to check for any discrepancies. Any significant differences would place sources in different redshift or $M_*$ bins, thereby changing the inferred $f_\mathrm{LERG}$ for the two methods of deriving physical properties. This check yielded similar distributions, with the medians and uncertainties close to each other.

\begin{figure}[htbp!]
    \centering
    \includegraphics[width=0.9\linewidth]{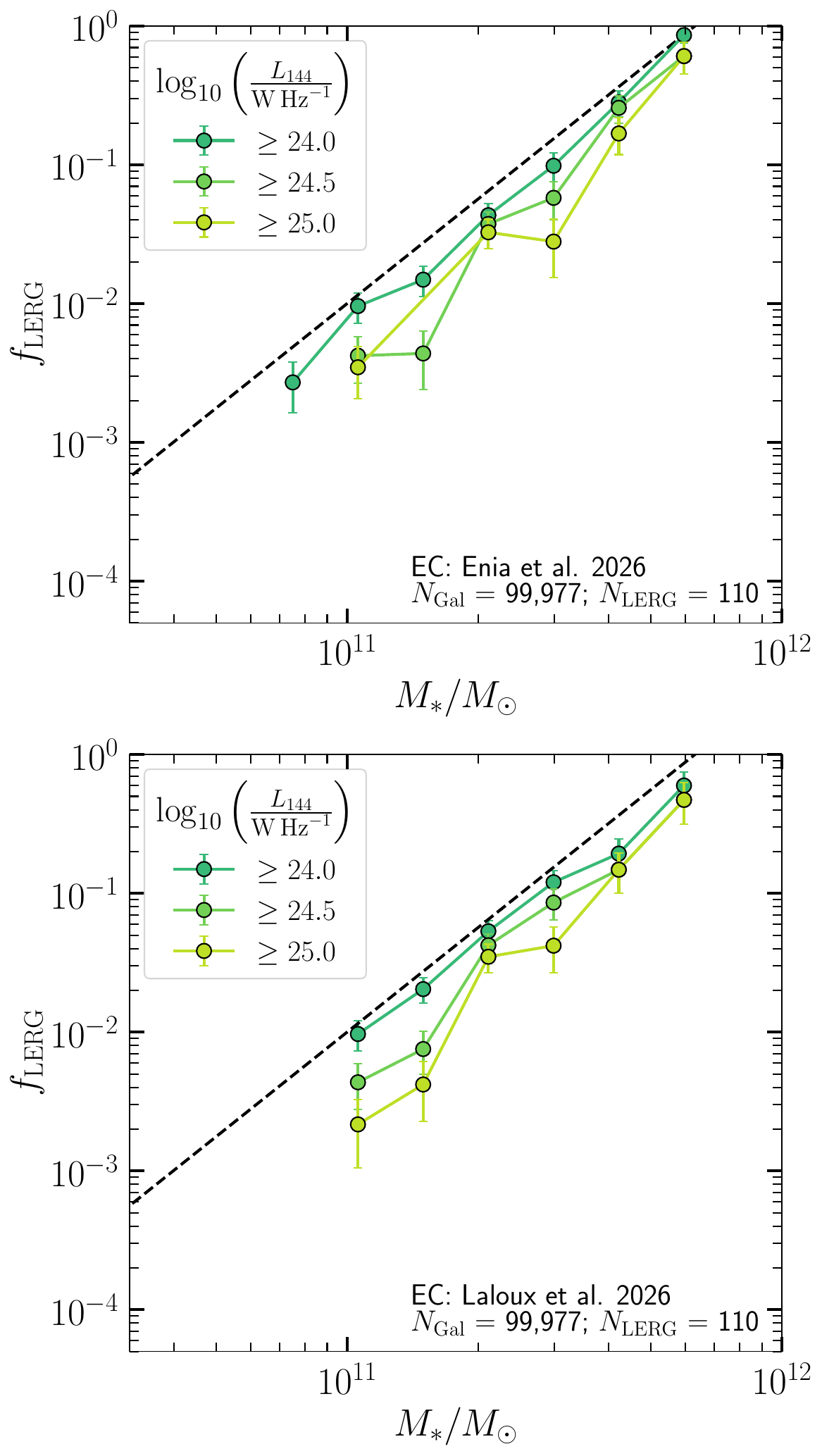}
    \caption{Same as in Fig.~\ref{fig:AGN fractions}, but only for the range \mbox{$0.7\leq z \leq 0.9$}. The black line in each panel shows the \mbox{$f_\mathrm{RLAGN}= 0.01\,(M_*/10^{11}M_\odot)^{2.5}$} relation found in the local Universe \citep{Best2005a}. \emph{Top}: Fraction $f_\mathrm{LERG}$ for the physical parameters derived in \citetalias{Q1-SP031}. \emph{Bottom}: Fraction $f_\mathrm{LERG}$ for the parameters derived in \citetalias{Q1-Laloux}.}
    \label{fig:compare_pp_fractions}
\end{figure}

\begin{figure}[htbp!]
    \centering
    \includegraphics[width=0.9\linewidth]{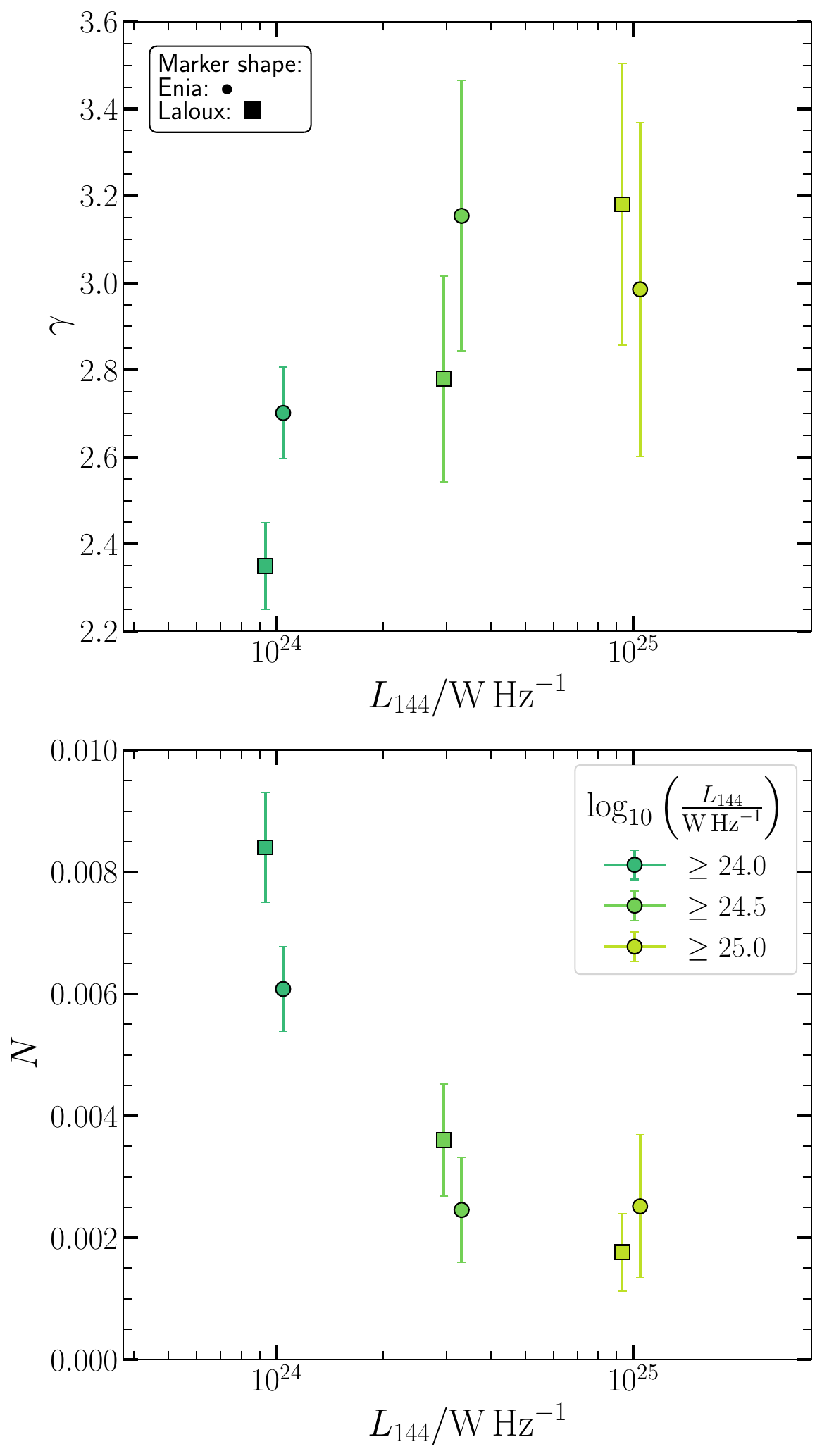}
    \caption{Fitted power-law parameters for different $L_{\rm 144}$ limits in the range \mbox{$0.7\leq z \leq 0.9$}, where different colours indicate the $L_{\rm 144}$ limits. \emph{Top}: Power-law index, where circles represent the parameters found for the physical parameters derived in \citetalias{Q1-SP031} and the squares show the results for the parameters derived in \citetalias{Q1-Laloux} for the same objects. \emph{Bottom}: Same as in the top panel, for the normalisation.}
    \label{fig:compare_fits_pp_fractions}
\end{figure}

We also calculate $f_\mathrm{LERG}$ as a function of $M_*$ and $L_{\rm 144}$ in the redshift range $0.7\leq z \leq0.9$ for physical parameters from both catalogues. This redshift range was chosen because the number of sources with physical parameters in both catalogues was largest. Within this redshift range, there are $148$ \acp{lerg} with physical parameters derived in \citetalias{Q1-Laloux}. For the same objects, we obtained photometric redshifts and $M_*$ from the catalogue in \citetalias{Q1-SP031} and calculated $f_\mathrm{LERG}$ for both, as seen in Fig.~\ref{fig:compare_pp_fractions}. Note that this results in slightly fewer objects, $110$, with physical parameters in both catalogues, since we required reliable parameters for both catalogues and that the photometric redshifts lie in the same redshift range. Subsequently, we fitted the resulting $f_\mathrm{LERG}$ with a power law (see Eq.~\ref{eq:powerlaw}) to assess consistency. For the sources with parameters derived in \citetalias{Q1-Laloux}, we see a mildly higher normalisation. However, the fitted power-law parameters shown in Fig.~\ref{fig:compare_fits_pp_fractions} were found to be consistent within uncertainties. This shows that even with a quite limited number of objects, both methods agree, and this test confirms the reliability of the results discussed before.

\begin{figure*}[t]
    \centering
    \includegraphics[width=\linewidth]{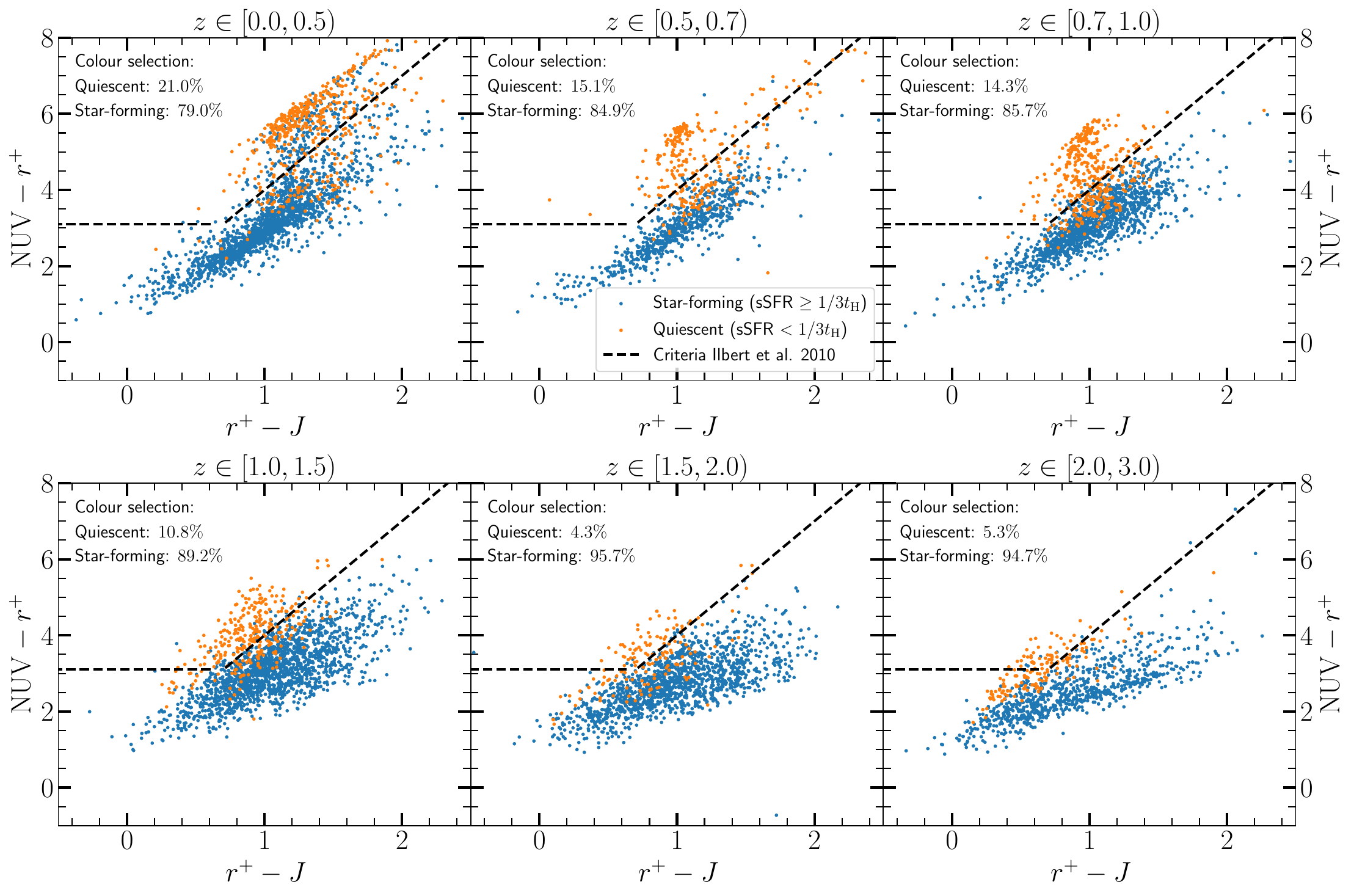}
    \caption{$\rm NUV$--$r^+$ and $r^+$--$J$ colours for all radio sources with a \Euclid match in our final sample. Using the colour criteria defined by \cite{Ilbert2010GalaxySurvey}, we classify galaxies into star-forming and quiescent and compare this to the same classification using the \ac{ssfr} for different redshift bins. For the classification using the \ac{ssfr}, we show sources classified as star-forming in blue and sources classified as quiescent in orange.}
    \label{fig:nuv-r-j colours}
\end{figure*}

\section{Selecting quiescent and star-forming galaxies}\label{app: quiet/sf galaxies}
As discussed in Sect.~\ref{sub-sc:SF vs quiescent}, we split our sample of \acp{lerg} into quiescent and star-forming galaxies using the \ac{ssfr} to investigate if these two populations evolve differently with $z$ and $M_*$. To check our selection method, we use \mbox{$\rm NUV$--$r^+$} and \mbox{$r^+$--$J$} colours to select quiescent galaxies that satisfy (\citealt{Ilbert2010GalaxySurvey}; \citetalias{Q1-SP031})
\begin{equation} \label{eq:Ilbert_colours}
    \begin{aligned}
    & \mathrm{NUV}-r^{+}>3\left(r^{+}-J\right)+1\; ; \\
    & \mathrm{NUV}-r^{+}>3.1\; .
    \end{aligned}
\end{equation}
This combination of criteria is analogous to the $UVJ$ diagram. It has become widely used in recent studies because the \mbox{$\rm NUV$--$r^+$} colour is more sensitive to recent star formation, providing a cleaner separation between passive (redder) and star-forming (bluer) galaxies \citep{Arnouts2013EncodingGalaxies, Ilbert2013MassUltraVISTA}. 

In Fig.~\ref{fig:nuv-r-j colours}, we compare the classification of galaxies as quiescent or star-forming based on the \ac{ssfr} with that obtained using the colour criteria defined in Eq.~\ref{eq:Ilbert_colours}. Since we require a detection in all three bands (NUV, $r^+$, $J$), the comparison includes $9\,815$ out of $10\,794$ radio sources with a \Euclid counterpart from the final sample (Sect.~\ref{sub-sc:Final sample}). Figure~\ref{fig:nuv-r-j colours} shows good agreement between the two classification methods, indicating that the adopted \ac{ssfr}-based classification is robust. Applying the colour selection to the $2\,656$ detected \acp{lerg} yields $864$ quiescent ($32.5\%$) and $1\,792$ star-forming ($65.5\%$) sources, meaning that the colour selection yields relatively more star-forming objects. Although a more stringent threshold of $f_{\rm sSFR}=1/5$ would produce closer agreement with the colour selection, we adopt $f_{\rm sSFR}=1/3$ in Sect.~\ref{sub-sc:SF vs quiescent} to be consistent with \cite{Kondapally2022, Kondapally2025}. We verified that this choice does not significantly affect our results by calculating the \ac{lerg} incidence for both thresholds.

\section{Fitting fractions of star-forming and quiescent LERGs}\label{app: quiescent and sf lerg fractions}
\begin{figure*}[t]
    \centering
    \includegraphics[width=\linewidth]{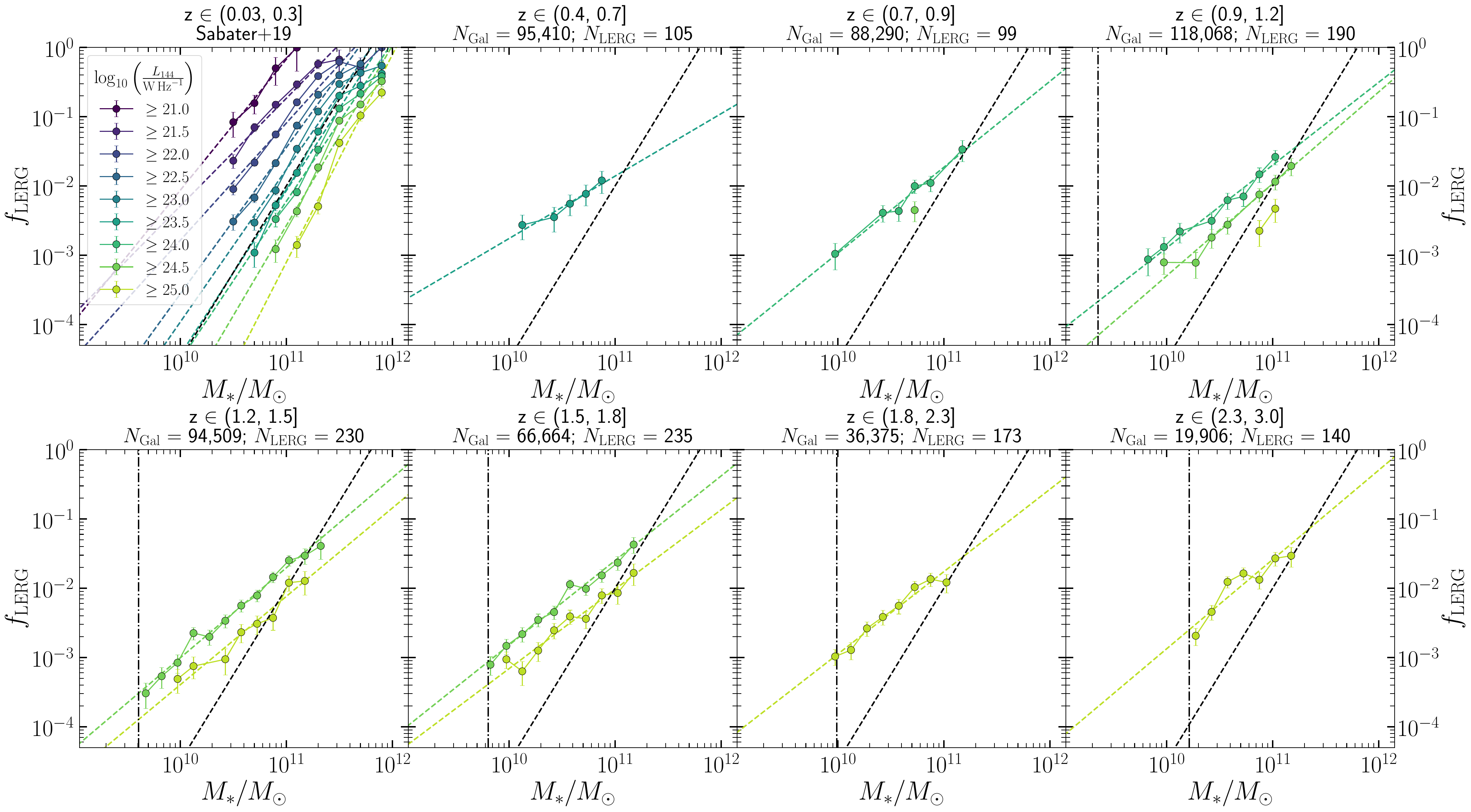}
    \caption{Same as in Fig.~\ref{fig:AGN fractions}, but for star-forming \acp{lerg}.}
    \label{fig:fitted star-forming AGN fractions}
\end{figure*}

\begin{figure*}[t]
    \centering
    \includegraphics[width=\linewidth]{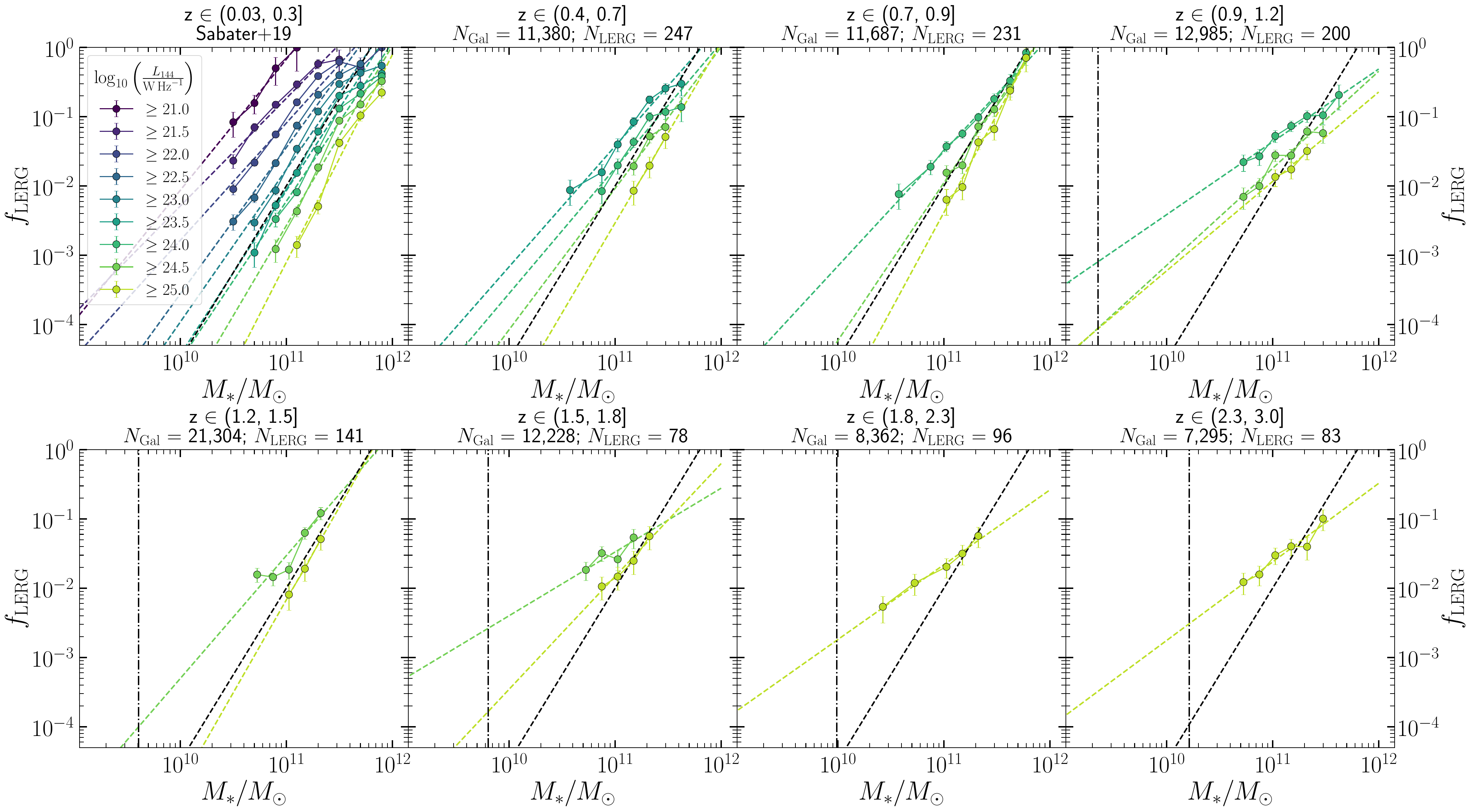}
    \caption{Same as in Fig.~\ref{fig:AGN fractions}, but for quiescent \acp{lerg}.}
    \label{fig:fitted quiescent AGN fractions}
\end{figure*}

As discussed in Sect.~\ref{sub-sc:SF vs quiescent}, we separated all \acp{lerg} into star-forming and quiescent galaxies and calculated $f_\mathrm{LERG}$ for each class. Subsequently, we used \ac{odr} to fit the calculated fractions $f_\mathrm{LERG}$ for each $L_{\rm 144}$ limit in all redshift bins that have at least three $f_\mathrm{LERG}$ values. The fraction $f_\mathrm{LERG}$ for star-forming and quiescent \acp{lerg} and the best power-law fits for every $L_{144}$ limit in all redshift bins are shown in Figs.~\ref{fig:fitted star-forming AGN fractions} and \ref{fig:fitted quiescent AGN fractions}, respectively.

\end{appendix}

\end{document}